\documentclass[11pt,a4paper]{article}
\pdfoutput=1
\usepackage[utf8]{inputenc}
\usepackage{jheppub, slashed, multirow, xcolor}
\usepackage[normalem]{ulem}

\makeatletter\g@addto@macro\bfseries{\boldmath}

\newcommand{\app}[1]{\hyperref[#1]{Appendix~\ref{#1}}}
\def\equationautorefname~#1\null{Eq.\,(#1)\null}

\def\be   {\begin{equation}}   \def\ee   {\end{equation}}
\def\bea  {\begin{eqnarray}}   \def\eea  {\end{eqnarray}}
\def\nn{\nonumber}

\newcommand{\descriptionmargin}{5mm}

\preprint{
\begin{flushright}
DESY 17-049\\
IFIC/17-15\\
FTUV-17-0322.9647\\
\end{flushright}}

\title{Minimally extended SILH}

\author[a,b]{Mikael Chala,}		
\author[a]{Gauthier Durieux,}	
\author[,a,c]{Christophe Grojean,\footnote{On leave from Instituci\'o Catalana de Recerca i Estudis Avan\c cats, 08010 Barcelona, Spain.}}	
\author[a,d]{Leonardo de Lima}	
\author[a]{and Oleksii Matsedonskyi}	

\affiliation[a]{DESY, Notkestra\ss e 85, D-22607 Hamburg, Germany}
\affiliation[b]{Departament de F\'isica T\`eorica, Universitat de Val\`encia and IFIC, Universitat de Val\`encia-CSIC, Dr. Moliner 50, E-46100 Burjassot (Val\`encia), Spain}
\affiliation[c]{Institut f\"ur Physik, Humboldt-Universit\"at zu Berlin, D-12489 Berlin, Germany}
\affiliation[d]{Instituto de F\'isica Te\'orica, Universidade Estadual Paulista, SP, Brazil}

\emailAdd{mikael.chala@ific.uv.es}
\emailAdd{leolima@ift.unesp.br}
\emailAdd{gauthier.durieux@desy.de}
\emailAdd{christophe.grojean@desy.de}
\emailAdd{oleksii.matsedonskyi@desy.de}

\abstract{Higgs boson compositeness is a phenomenologically viable scenario addressing the hierarchy problem. In minimal models, the Higgs boson is the only degree of freedom of the strong sector below the strong interaction scale. We present here the simplest extension of such a framework with an additional composite spin-zero singlet.
To this end, we adopt an effective field theory approach and develop a set of rules to estimate the size of the various operator coefficients, relating them to the parameters of the strong sector and its structural features. As a result, we obtain the patterns of new interactions affecting both the new singlet and the Higgs boson's physics. We identify the characteristics of the singlet field which cause its effects on Higgs physics to dominate over the ones inherited from the composite nature of the Higgs boson. Our effective field theory construction is supported by comparisons with explicit UV models.
}

\begin{document}

\maketitle


\section{Introduction}

Here and there, in collider experiments, we see hints of deviations from the standard model (SM) and will probably see more in the future. These anomalies ---e.g., in the diphoton spectrum, weak diboson spectrum, or in the flavor sector--- may point at new physics in close vicinity of the electroweak (EW) scale which is well motivated theoretically given the gauge hierarchy problem. One of the most popular solutions to this puzzle is provided by the Higgs compositeness paradigm~\cite{ch,Agashe:2004rs}\footnote{For detailed reviews of composite PNGB Higgs we for instance refer the reader to Refs.~\cite{Contino:2010rs,Panico:2015jxa,Bellazzini:2014yua}.} whose implications at EW energies can be reflected in an economical and reasonably model-independent way in the strongly interacting light Higgs (SILH) framework~\cite{Giudice:2007fh}. It does not only address the construction of a basis of dimension-six operators from standard-model fields, but also imposes the theoretical biases arising from our understanding of strongly coupled ultraviolet (UV) completions. Its predictivity is thereby increased. This framework can in principle be extended to also describe new states. In this note, we aim at presenting the minimal such extension including one single new dynamical degree of freedom in the form of a composite spin-zero EW singlet $S$, heavier than the Higgs boson. There are numerous roles an extra singlet field could play and help addressing relevant issues that the standard model is deficient with, like the abundance of dark matter~(see e.g. Refs.~\cite{Frigerio:2012uc,Bruggisser:2016ixa}) or the matter-antimatter asymmetry that could be produced during a strong first order electroweak phase transition~\cite{Espinosa:2011eu, vonHarling:2016vhf}.

Since its discovery, the Higgs boson experienced a change of status and has become a tool to search for new physics. In composite Higgs (CH) models, the deviations from the SM predictions of the Higgs production and decay rates induced by the putative strong dynamics above the EW scale have been known for quite some time. The goal of our study is to understand how these generic signatures are affected by the presence of additional composite states below the new scale of the strong interactions. In particular, we identify under which circumstances the effects of a singlet spin-zero field dominate over that of the strong sector, in which channels they are more likely to be revealed and how they can help deciphering the dynamics governing this singlet field.

The structure of this paper is the following. In \autoref{sec:SUV}, we define the general power counting rule for the operators involving SM fields and $S$, discuss possible underlying UV dynamics, and construct a basis from the relevant operators. In \autoref{sec:HSscen}, we apply our formalism to several minimal but consistent scenarios for the composite $H+S$ pair and discuss their phenomenology, notably in the Higgs sector. In \autoref{sec:specmod}, we discuss the matching of two explicit examples of composite Higgs models onto our power counting. Finally, our conclusions are presented in \autoref{sec:conc}.

\section{General formalism}
\label{sec:SUV}

In this section, we set up an effective field theory (EFT) framework describing a variety of models that feature a new strongly coupled dynamics which confines at some scale $f$ not very far above the EW scale. We assume that the Higgs doublet $H$ and a spin-zero gauge singlet $S$ are the lightest composite resonances and include them explicitly in our EFT. The effects of the rest of the strong dynamics are described by effective operators. This type of spectrum can naturally arise in theories where both $S$ and the Higgs are pseudo Nambu--Goldstone bosons (PNGB) associated with the spontaneous breaking of an approximate global symmetry of the strong sector at the scale $f$ (see Refs.~\cite{Gripaios:2009pe, Belyaev:2015hgo} for specific examples).
The general mechanism can be illustrated using the $SO(6) \to SO(5)$ spontaneous symmetry breaking pattern~\cite{Gripaios:2009pe}. It gives rise to five PNGBs transforming in a fundamental representation of $SO(5)$. Among those five states, a quadruplet and a singlet of $SO(4)\subset SO(5)$ are found. The quadruplet has the right quantum numbers to form a complex Higgs doublet if the SM EW group $SU(2)_L$ is identified with one of the factors of $SO(4) \sim SU(2)_1 \times SU(2)_2$, and the remaining $SO(4)$ singlet becomes the additional singlet $S$.
While assuming a PNGB nature for the Higgs seems indispensable,\footnote{For a brief analysis of non-PNGB, accidentally light Higgs we refer the reader to Ref.~\cite{Liu:2016idz}.} the relative lightness of $S$ may just be accidental. This situation can be realized for instance in scenarios with $SO(5) \to SO(4)$ breaking, giving only rise to four PNGBs forming the Higgs doublet.

Our aim here is to probe the robustness of the generic predictions of the minimal models and to construct a unified framework suitable for describing next-to-minimal models, thus capturing a broad range of explicit models and providing a common basis for their comparison.
In the following, we establish a general procedure to construct the effective Lagrangian of these next-to-minimal theories. In \autoref{sec:HSscen}, we use this procedure to build several consistent realizations of composite $S+H$ scenarios, corresponding to typical limiting cases.

\subsection{Power counting rule}
\label{sec:powcount}

Our EFT is complemented by a power counting rule allowing to estimate, up to order-one factors, the dependence of different operators on the strong sector properties. We assume that the quantitative features of the strong dynamics can be fully characterized by a typical mass $m_\rho$ of the lightest composite states other than $S$ and $H$, playing also the role of a cutoff of our EFT description, and a typical coupling $g_\rho$, which can take values in a range from order-one to $4\pi$. The expected scaling of the effective operator coefficients in terms of these parameters can be determined using dimensional analysis.
The qualitative features of the strong sector, related to its symmetries and internal structure, manifest themselves through selection rules, i.e. the suppression of certain operator coefficients with respect to generic expectations.
In the reminder of this section, after having introduced the required dimensional analysis arguments, we discuss the selection rules deriving from the nature of $S$ and the possible features of an underlying UV completion. We cover scenarios featuring a CP-odd or -even, generic or PNGB, $S$ state.

\subsubsection*{Dimensional analysis}

With the $c=\hbar=1$ convention abandoned, dimensional analysis determines the correct scaling of the EFT operators coefficients with $m_\rho$ and $g_\rho$. Following Ref.~\cite{Panico:2015jxa}, we characterize the dimensions of all the relevant objects in units of $\hbar$ and length $L$. The $\hbar$ and $L$ dimensions of an operator generated at $\#_L$ loops are $(1-\#_L)$ and $-4$ respectively. They have to match the total dimension of the scalars ($\hbar$ dimension 1/2, $L$ dimension $-1$), fermions (1/2, and $-3/2$), vectors (1/2 and $-1$), mass parameters, derivatives (0 and $-1$) and couplings ($-1/2$ and 0) it involves.
Hence, if no selection rule applies, an operator with $\#_H$ external fields $H$, $\#_S$ external $S$, and $\#_{\partial}$ derivatives has the form
\be \label{eq:powc_1}
{m_\rho^4 \over g_\rho^2}  \left[{g_\rho S \over m_\rho}\right]^{\#_S} \left[{g_\rho H \over m_\rho}\right]^{\#_H} \left[{\partial_\mu \over m_\rho}\right]^{\#_{\partial}}
\qquad\text{(composite states)}\,.
\ee
Selection rules can however play a crucial role in determining the magnitude of operator coefficients. For instance, CP invariance in the strong sector would forbid the $S|H|^2$ and $S|D_\mu H|^2$ operators for a pseudo-scalar $S$. Less trivial examples will be considered in the following.

According to the rule~\eqref{eq:powc_1}, the insertion of a Higgs field is associated with a factor $g_\rho / m_\rho$. Comparing this to the usual parametrization of Goldstone bosons, appearing in the Lagrangian only through $U\sim \exp(i H/f)$, we obtain the important relation:
\be\label{eq:mrho}
m_\rho \sim  g_\rho f \,.
\ee
It is worth emphasizing that $f$ and $m_\rho$ thus have different $\hbar$ dimensions.

\subsubsection*{Shift symmetry breaking and partial compositeness}

An important symmetry encountered in CH models is the ``shift" symmetry of the Nambu--Goldstone bosons. If unbroken, it forbids Goldstone bosons to have a potential and, in particular, a mass.
In realistic models, the PNGB Higgs potential is generated through the partial compositeness (PC) mechanism.\footnote{Partial compositeness of the top quark seems to be the only viable way to make the top as heavy as it is, while the mass of other SM states can in principle be generated in a different way, see Refs.~\cite{Matsedonskyi:2014iha,Cacciapaglia:2015dsa,Panico:2016ull}. See also Ref.~\cite{Galloway:2010bp} for a CH model with other sources of shift symmetry breaking.} In this case, the breaking of the shift symmetry is induced by the couplings of the strong sector to the rest of the SM fields, which are assumed to be elementary.
It is rather natural to suppose that $S$, if realized as a PNGB, shares its shift symmetry breaking source with the Higgs boson.\footnote{Though the Higgs and $S$ shift symmetries and breaking sources are {\it a priori} independent, we take them to be equal as a first approximation, keeping in mind that this assumption can be relaxed.}
Since the shift-breaking interactions couple elementary SM states to the Higgs boson, they are also responsible for the generation of SM masses.
The couplings to the heaviest SM fermion, namely the top quark, hence induce the largest breaking. Therefore, all the shift symmetry breaking operators have to either explicitly contain SM fields or be suppressed by a loop of the elementary top quark. This loop suppression can be estimated from dimensional arguments as $N_c y_t^2/(4\pi)^2$ or $N_c y_t g_\rho/(4\pi)^2$, where $y_t$ is the SM top quark Yukawa coupling and $N_c$ is a number of colors. Either value can appear in explicit models~\cite{Panico:2012uw}. In the following, we will stick to the first option which gives the most distinct results with respect to the non-PNGB case.
This discussion can be formalized by adding the following factors to the power counting formula~(\ref{eq:powc_1}):
\be \label{eq:powc_2}
\left[{ N_c y_t^2 \over (4 \pi)^2}\right]^{\#_{\slashed L}}   \left[{y_q \bar q q \over m_\rho^{2} f}\right]^{\#_{\bar q q}}
\left[{g_A A \over m_\rho}\right]^{\#_A}
\qquad\text{(elementary states)}\, .
\ee
Here, $y_q$ is the SM Yukawa coupling of the fermion $q$, $g_A$ is the coupling strength of the SM gauge field $A$, and $\#_{\bar q q}, \#_{A}, \#_{\slashed L}$ are respectively the numbers of fermion bilinears, gauge fields, and loop suppression factors required to break the $S$ or $H$ shift symmetry.
The parametric form of the coefficient again follows from dimensional analysis, while the presence of a Yukawa coupling in front of the fermion bilinear assumes Minimal Flavor Violation~\cite{DAmbrosio:2002vsn}. Note that it is in any case required for the Higgs interactions to reproduce  the form of the SM Yukawa interactions $y_q \bar q H q$, and that the $S$ couplings to fermions can be naturally endowed with this flavor structure, because they always involve a Higgs doublet. For simplicity we will assume that the chirality conserving quark bilinears of the type $\bar q \gamma_\mu q$ obey the same power counting as $\bar q q$.

Using the formulas~(\ref{eq:powc_1}),(\ref{eq:powc_2}) we can for example obtain the parametric form of the one-loop PNGB Higgs potential
\be\label{eq:Vh}
V_h = m_\rho^2 f^2 {N_c y_t^2 \over (4 \pi)^2} \left(-\alpha {|H|^2\over f^2}+\beta{|H|^4\over f^4}\right)
\ee
where $\alpha$ and $\beta$ are dimensionless coefficients which are expected to be of order one. Its minimization yields
\be
v = f \left({\alpha \over 2 \beta}\right)^{1\over2} \;\;\; \text{and} \;\;\; m_h^2 \simeq \beta {N_c y_t^2 \over 2 \pi^2}  \, {v^2 \over f^2} \, m_\rho^2\, ,
\ee
where $m_h$ is the Higgs mass and $v$ its vacuum expectation value (VEV).\footnote{Notice that the Higgs field value is not proportional to the symmetry breaking parameters because, in the absence of external breaking, there is no Higgs potential, the Higgs VEV is simply not fixed and can take any value, i.e. one should not expect that $v\to0$ for $y_t\to0$.} A key parameter of CH models is the ratio $\xi = {v^2 / f^2}$ of the EW symmetry breaking scale $v\sim246$\,GeV and the strong sector global symmetry breaking scale $f$. It controls the size of the Higgs couplings deformations with respect to SM predictions~\cite{Azatov:2011qy, Contino:2013kra} and is already bounded to be $\xi\lesssim 0.2$~\cite{Grojean:2013qca, Matsedonskyi:2015dns}. In order to achieve the separation $v\ll f$ required phenomenologically, one has to tune the $\alpha$ and $\beta$ coefficients of the potential. An additional tuning of the $\beta$ coefficient may be required to provide a sufficiently low Higgs mass. If $S$ is a PNGB as well, its potential would have the same parametric form as the Higgs, but there is {\it a priori} no reason for any tuning to take place. One therefore expects the following hierarchy between the Higgs boson mass, the $S$ mass $M$, and the masses of other composite states
\be\;\label{eq:massrelations}
m_h^2 : M^2 : m_\rho^2 \;\sim \; {N_c y_t^2 \over (4\pi)^2} \xi : {N_c y_t^2 \over (4\pi)^2} : 1 \qquad\text{(PNGB S with PC breaking)}\,.
\ee
In the case of a generic $S$, we rather expect this mass hierarchy to be
\be
m_h^2 : M^2 : m_\rho^2 \; \sim \; {N_c y_t^2 \over (4\pi)^2} \xi : 1 : 1
\qquad\text{(generic S)}\,.
\ee
The EFT validity then requires $S$ to be accidentally lighter than the cutoff $m_\rho$, with the degree of tuning $M^2/m_\rho^2$ characterizing the accuracy of our description.

\subsubsection*{Anomaly-mediated shift symmetry breaking}

A breaking of the shift symmetry of a PNGB $S$ through PC is not the only possibility. It is actually not strictly necessary since $S$ does not have to couple to SM fermions, unlike $H$ which generates their masses. The shift symmetry of a CP-odd $S$ can for instance be broken by anomalies associated with the gauge fields of the SM or the strong sector
\be \label{eq:anom1}
{N_f g_X^2 \over (4 \pi)^2}  {S\over f} X_{\mu \nu} \tilde X^{\mu \nu} \,,
\ee
where $X_{\mu \nu}$ is a gauge field strength tensor, $g_X$ the corresponding coupling and $N_f$ is an anomaly coefficient roughly corresponding to the number of strong sector fermion flavors generating the anomaly. The anomalous interactions with the SM gauge fields can however not generate a sufficiently large $S$ mass~\cite{Bellazzini:2015nxw}. In order to make $S$ heavier than the Higgs boson, one could either again resort to PC breaking, or assume that $M$ arises from the anomaly related to the new strong dynamics. In the latter case, in analogy with the $\eta^\prime$ meson of QCD~\cite{eta}, we obtain $M^2 \simeq m_\rho^2\; N_f/N$, where $N$ is a number of colors of the underlying strong dynamics.
Using the relation $1/N \sim g_\rho^2 / (4 \pi)^2$, predicted for large-$N$ theories (see next section), the expression for the $S$ mass can be rewritten as $M^2 \simeq m_\rho^2 \; N_f g_\rho^2/(4 \pi)^2$. So, in this case, the estimate for the mass hierarchy is
\be
m_h^2 : M^2 : m_\rho^2 \; \sim \; {N_c y_t^2 \over (4\pi)^2} \xi : {N_f g_\rho^2 \over (4\pi)^2} : 1
\qquad\text{(PNGB S with anom. breaking)}\,.
\ee
The $S$ mass and couplings to gauge bosons differ from the generic estimates, $m_\rho^2 S^2$ and ${g_X^2 / g_\rho^2}$ $X_{\mu \nu} \tilde X^{\mu \nu} S/f$, derived from Eqs.~(\ref{eq:powc_1}),~(\ref{eq:powc_2}) by a factor of $N_f g_\rho^2 /(4 \pi)^2$, which we thus include as a suppression characteristic of anomaly breaking in our power counting rule.
Notice that the $N_f$ factors appearing in the anomalous couplings and in the expression for the mass are in general independent.

\subsubsection*{UV selection rules}

We have so far discussed selection rules connected to symmetry breaking.
In addition, some of the operators can carry suppressions not transparently related to the EFT symmetries.
Two types of such suppressions, present in large-$N$ and $N$-site theories, will be described in the following two sections. They can affect the couplings of $S$ to the SM gauge bosons or the Higgs field, leading to an additional ${N_f g_\rho^2 / (4 \pi)^2}$ factor, where $N_f$ is an effective number of composite flavors.

At this point we can summarize the power counting in a single expression
\be \label{eq:nda}
m_\rho^2 f^2  \left[{ N_c y_t^2 \over (4 \pi)^2}\right]^{\#_{\slashed L}} \left[{ N_f g_\rho^2 \over (4 \pi)^2}\right]^{\#_{L}}   \left[{y_q \bar q q \over m_\rho^{2} f}\right]^{\#_{\bar q q}}
\left[{g_A A \over m_\rho}\right]^{\#_A} \left[{S \over f}\right]^{\#_S} \left[{H \over f}\right]^{\#_H} \left[{\partial_\mu \over m_\rho}\right]^{\#_\partial} \,,
\ee
where $\#_{\slashed L}$ is a number of loops required to break the shift symmetry through PC, $\#_{L}$ stands for a number of loops required by the UV selection rules or the shift symmetry breaking by anomalies, and the remaining $\#$'s correspond to the number of insertions of external fields or momenta. The power counting formula~(\ref{eq:nda}) applies only to the operators generated by the strong dynamics and, for instance, not to the elementary field kinetic terms. It complements the power countings developed in Refs.~\cite{Giudice:2007fh,Liu:2016idz,Panico:2011pw} for CH models, in what concerns the presence of an additional state $S$, but does not have their generality, as we made simplifications to display more transparently the physics relevant for our discussion.

Having defined the basic ingredients of our EFT, we now comment on its validity. As usual, we have to limit our EFT description to operators of a certain mass dimension. In order to keep the effect of higher-dimensional operators negligible, we need a sizable separation between $M$ and $m_\rho$. As we have seen, a PNGB $S$ can be parametrically lighter than $m_\rho$, while for a generic $S$ the scale separation could be accidental or due to some unknown features of the underlying strong dynamics leading to deviations from our power counting estimates.
However it is worth stressing that the first signals of a new resonance will not allow for a precise determination of its properties. Instead, one will only be  sensitive to the order of magnitude of different operator coefficients and therefore to selection rules. Hence, even with a moderate $M-m_\rho$ separation, our framework could allow to determine the main features of the underlying theory and could point at the explicit UV completions of the most appropriate type.

In the two following subsections we give a short overview of the two well-known approaches used to describe the behavior of strongly coupled dynamics bound states. They lead to ---and provide us with further insight in--- the power counting rule~(\ref{eq:nda}).

\subsection{Matching to large-\texorpdfstring{$N$}{N} theories}
\label{sec:match}

As a first prototypical example of UV completion we consider confining $SU(N)$ gauge theories with $N_f$ quark flavors transforming in the fundamental representation of the gauge group. (We will call the new states \emph{quarks} and \emph{gluons} for simplicity and will not refer to their SM analogues in this section.) In this case, we can use a $1/N$ expansion~\cite{'tHooft:1973jz,'tHooft:1974hx,Witten:1979kh} in order to understand the properties of bound states.
The bound states which we are interested in form when the coupling  $g_*$ between quarks and gluons becomes strong, hence an expansion in $g_*$ is not useful for their description. But in the strongly coupled regime characterized by
\be\label{eq:largeN}
N {g_*^2\over 16 \pi^2} \sim 1\,,
\ee
amplitudes acquire a well-defined scaling with $N$ which can be used to estimate their relative size.
The use of the $1/N$ expansion  relies on the assumption that this regime plays the dominant role in the bound state dynamics.

Let us consider the specific example of meson-like states which are typically the lightest and therefore can be good candidates for $H$ and $S$. In the Feynman diagrams corresponding to meson interactions, each additional gluon loop brings an extra factor of $N {g_*^2/ 16 \pi^2}\sim 1$. This means that diagrams with any number of additional gluon propagator insertions have at most\footnote{The largest contributions are given by the planar diagrams, while the non-planar ones carry extra $1/N$ suppression factors.} the same size as the leading-order diagram. Their sum therefore has the same scaling with $N$ as the easily estimated leading-order contribution. Using this feature, one can determine the expected scaling of different $n$-point functions~\cite{'tHooft:1973jz, 'tHooft:1974hx, Witten:1979kh}, effectively resuming an infinite series in $g_*$. The power counting formula (\ref{eq:powc_1}) is then recovered for the interaction of meson-like states with the following identification for the meson-meson coupling strength
\be \label{eq:grho}
g_\rho = {4 \pi \over \sqrt N}\, .
\ee
One can also show that the mass of mesons $m_\rho$ is independent of $N$~\cite{Witten:1979kh}. Notice that glueballs and baryons behave differently: our power counting only applies to mesons.

\begin{figure}
\centering
\includegraphics[width=.25\textwidth]{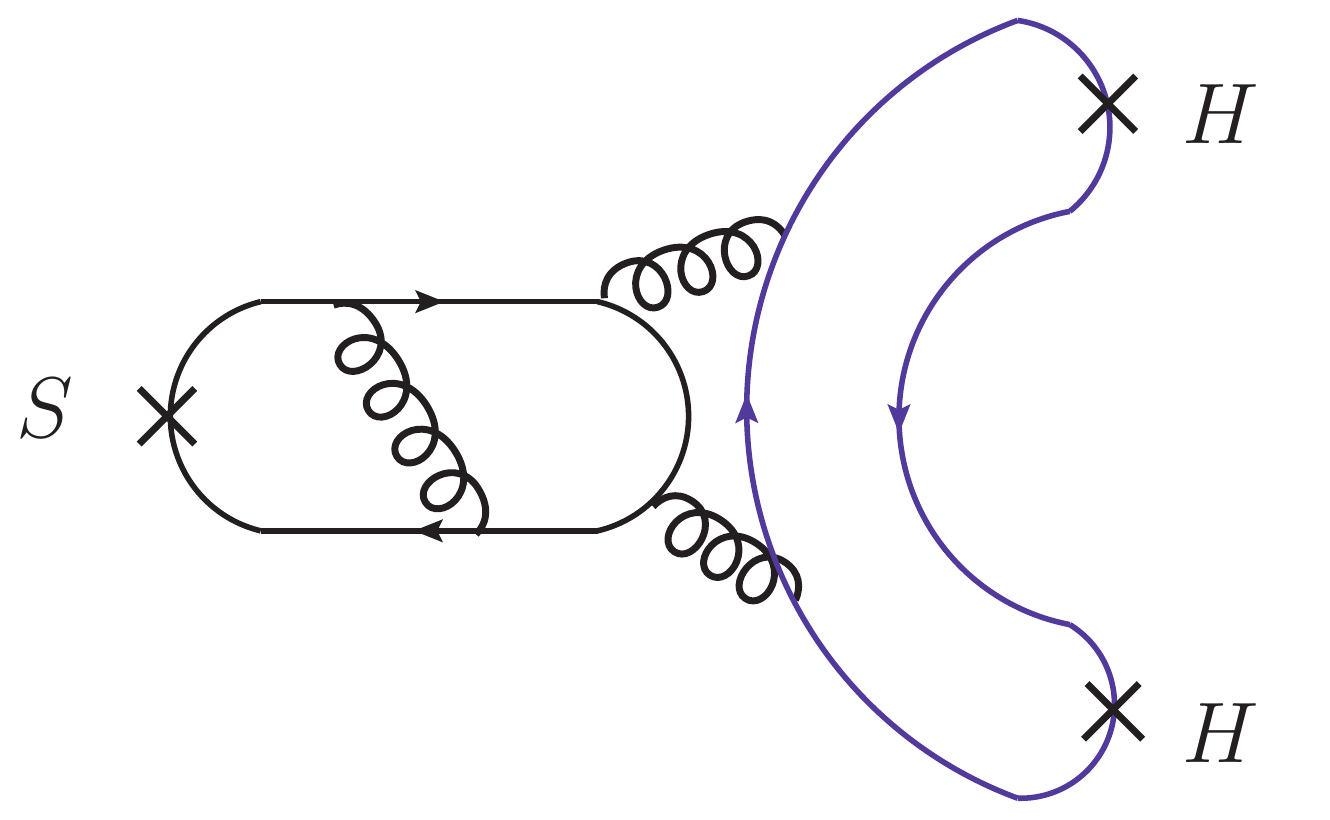}
\hspace{1cm}
\includegraphics[width=.5\textwidth]{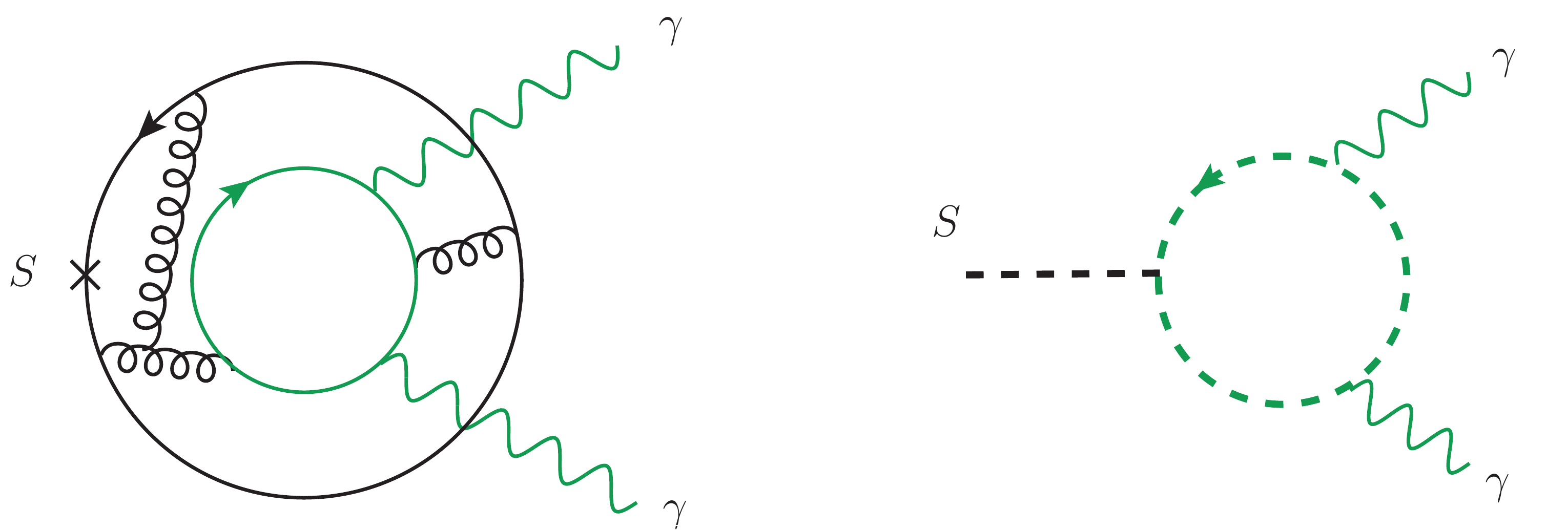}
\caption{Examples of diagrams generating $S|H|^2$ and $S|D_\mu H|^2$ couplings (left) and  $SF_{\mu\nu}F^{\mu\nu}$ coupling at loop level in terms of hypothetical hyperquark constituents of $S$ (center) and in terms of their bound states (right). Solid straight lines of different colors correspond to different fermionic flavors, wavy lines correspond to SM gauge bosons, and dashed lines are for composite scalars.}
\label{fig:graphs}
\end{figure}

Moreover, each additional quark loop brings an extra $1/N$ suppression. This derives from the fact that quarks have one fewer color index than gluons whose loops are unsuppressed when $N{g_*^2/ 16 \pi^2} \sim 1$.
Given the identification~(\ref{eq:grho}), a $1/N$ factor corresponds to a $g_\rho^2/ 16 \pi^2$ suppression, i.e., to a loop factor in the (\ref{eq:nda}) counting. The Zweig rule is an example of such suppression at work in QCD. Analogous suppressions can also appear in the interactions of composite states made of different types of quarks. With a different quark composition for $S$ and $H$, the $S |H|^2$ operator would for instance arise from a diagram containing two closed fermion lines instead of one for operators like $S^3$ featuring one single type of meson (see left graph of \autoref{fig:graphs}).

Closed fermionic lines can however be enhanced by the quark multiplicity, hence a factor $N_f$ in \autoref{eq:nda}. At the same time the scaling of meson masses and couplings with $N$ is not affected by $N_f$ as long as one remains within the region of applicability of the large-$N$ expansion which requires $N_f<N$ and also ensures confinement~\cite{Luty:1994ua}. As an example, let us consider the coupling of $S$ to SM gauge field strengths if the quark constituents of $S$ are SM-neutral. It must involve a loop of other quarks, charged under SM, which brings a $1/N$ suppression together with an enhancement by the number of quark flavors running in this additional loop. This example is represented graphically on the central graph of \autoref{fig:graphs}. Note the analogy with a process induced by a loop of SM-charged mesons (right graph of \autoref{fig:graphs}).

\subsection{Matching to multisite models}
\label{sec:nsite}

Multisite models are often used as a weakly coupled description of the lowest laying composite resonances~\cite{Foadi:2010bu,Panico:2011pw,DeCurtis:2011yx}, inspired by five-dimensional realizations of the composite Higgs~\cite{ch} and the idea of dimensional deconstruction~\cite{ArkaniHamed:2001nc}. In this section, we give a general overview of the relation between the two-site models and the power counting rules developed in \autoref{sec:powcount}. Two concrete examples of two-site models will be discussed later in \autoref{sec:specmod}.

Two-site models consist of two separate sites in a theory space, each featuring separately a copy of an approximate global symmetry $G$, which we call $G_1$ and $G_2$, and containing certain sets of gauge and matter fields. The product $G_1 \times G_2$ is spontaneously broken to the diagonal subgroup $G_{\text{diag}}$. The Goldstone bosons $\chi$ of the spontaneous breaking are embedded into the unitary matrix $U=\exp[i \chi / f]$, which transforms under $G_1 \times G_2$ rotations as
\be
U \to g_1 U g_2^\dagger\, .
\ee
Once set to its VEV, $\langle U \rangle = \mathbb{I}$, the Goldstone matrix only leaves unbroken the subgroup $G_{\text{diag}}$, corresponding to transformations with $g_1=g_2$.
The field content of the first site is that of the SM without the Higgs, and the SM gauge symmetry group $G_{\text{SM}}$ is realized as a subgroup of $G_1$. Since $G_\text{SM} \subset G_1$, the SM fields form only incomplete multiplets of $G_1$ and hence break it. This mimics the explicit weak breaking of the strong sector global symmetry $G\to G_{\text{SM}}$ by elementary fields. The second site features a gauge symmetry $\mathcal{H}\subset G_2$ with typical gauge couplings $g_\rho \gg g_{\text{SM}}$. The second site therefore plays the role of strong sector with a spontaneous $G \to \mathcal{H}$ breaking. Since $\mathcal{H}$ is gauged and broken by a condensate $\langle U \rangle$, the gauge bosons of the second site $\rho_\mu$ acquire a mass $m_\rho \sim g_\rho f$. The $\chi$ components corresponding to the $\mathcal{H}$ generators can be absorbed by the $\mathcal{H}$ gauge bosons and disappear from the spectrum. The remaining Goldstone bosons, associated with the $G/\mathcal{H}$ coset, contain the Higgs field and possibly also $S$.

The crucial assumptions about the two-site model include, besides the choice of the $G$ and $\mathcal{H}$ symmetry groups, the embedding of the SM third-family quarks in incomplete multiplets of $G$. Once they are specified, the elementary fermions of the first site can be coupled to the composite fermionic partners $\psi$ of the second site
\be \label{eq:mix}
{\cal L}_{\text{mix}} = y_L f \, \bar q_L U \psi_R + y_R f \, \bar t_R U \psi_L + \text{h.c.} \,,
\ee
where $q_L=(t_L,b_L)$ and $t_R$ are SM quarks, embedded in some representation of $G$.\footnote{In this case, we have explicitly assumed that elementary fermions are embedded into fundamental representations of $G$ while, for other choices, one may need a different form of mixing, as will be seen in \autoref{sec:specmod}.} This Lagrangian realizes the PC paradigm and leads to the top quark Yukawa Lagrangian ${\cal L}_{\text{Yukawa}} \supset {y_L y_R f \over m_\psi} \bar q_L H t_R$ where $m_\psi$ is the typical mass of the composite fermions. At the same time, since the interactions~(\ref{eq:mix}) couple the Goldstones to the $G\to G_{\text{SM}}$ and $G \to \mathcal{H}$ breaking sources, they generate the loop-level scalar potential $V(y_L, y_R,m_\psi)$ for the Higgs boson and the singlet $S$.
The Goldstone symmetry-preserving interactions arise from the kinetic term of the $U$ field
\be \label{eq:kinU}
{\cal L}_{\text{kin},\chi} = {f^2 \over 2} \text{Tr}[D_\mu U (D^\mu U)^\dagger]\,,
\ee
where $D_\mu U = \partial_\mu U - i g_A A_\mu U + i g_\rho U \rho_\mu$. This Lagrangian also contains mixings between the elementary and composite gauge fields which make the SM gauge bosons partially composite and break the Goldstone shift symmetry.

The well-known structure described above, including the specific assumptions about the $G$ and $\mathcal{H}$ symmetry groups as well as the field content of the second site, determines the Lagrangian at the {\it renormalizable} level (thinking of $f U$ as a dimension-one scalar). Since the second site states are just an effective description of the lightest composite resonances, one however also needs to describe the effects of the rest of the strong sector through higher-order operators. In the following, we will simply assume that there are no other composite resonances up to the cutoff $\Lambda = 4 \pi f$. At this energy the derivative couplings between the Goldstone bosons become non-perturbative and the theory of bound states has to be substituted by some other description. Therefore, all the higher-order operators generated by the strong dynamics arise at the cutoff scale $\Lambda$ and we use the naive dimensional analysis (NDA) prescription~\cite{Manohar:1983md} to estimate their size. NDA predictions correspond to those of the power counting formula~(\ref{eq:nda}) in the limit of the maximal coupling $g_\rho \to 4\pi$ and hence $m_\rho \to \Lambda$. Our description of the composite resonances of the second site thus obviously requires a sufficiently large separation $m_\rho < \Lambda$ or, equivalently, $g_\rho < 4 \pi$.

First, it is trivial to see that the operators obtained by integrating out the composite resonances at a scale $m_\rho$, at tree level, from the renormalizable Lagrangian will automatically follow the~(\ref{eq:nda}) prescription.
Now let us discuss how the NDA-sized operators coming from the scale $\Lambda$ will affect the low-energy physics below the scale $m_\rho$, and how this effect can be captured by the power counting rule~(\ref{eq:nda}). Recall that the derivation of our formula~(\ref{eq:nda}), used to describe the physics below $m_\rho$, was based on the assumption that $m_\rho$ is the only UV scale of the problem, while now we have additional effects coming from $\Lambda$.
One may naively assume that the NDA coefficients of the relevant operators, like the Higgs mass, will be enhanced by factors of $\Lambda / m_\rho$ with respect to~(\ref{eq:nda}) predictions. This does however not happen because of the symmetry structure of the two-site model which requires the simultaneous presence of both $G_1\to G_{SM}$ and $G_2\to \mathcal{H}$ breaking sources to generate the PNGB potential. If any of the two sites were $G$-symmetric, the Goldstone matrix could be eliminated by a unitary $G$-rotation from the mixing Lagrangians~(\ref{eq:mix}), (\ref{eq:kinU}).
Since the $G_2\to \mathcal{H}$ breaking comes from the masses $\sim m_\rho$ and couplings $\sim g_\rho$ of the second site fields, the PNGB potential is suppressed by an additional factor of $m_\rho^2 / \Lambda^2$ with respect to the NDA estimate and thus follows the prediction of our power counting~(\ref{eq:nda}).\footnote{An analogous reasoning can be applied to the PNGB $S$ mass, but implementing the large mass separation $M \ll \Lambda$ for a generic $S$ may require some {\it ad hoc} assumptions.}
As for the irrelevant operators, the NDA predictions for their coefficients are suppressed by powers of $g_\rho / 4\pi$ with respect to the predictions of~(\ref{eq:nda}). This suppression is thus similar to the effect of what we call UV selection rules.

This leads to an important conclusion: in this particular realization of the strong sector, the UV selection rule suppression of some couplings arises if they can not be generated after integrating out, at tree level, the composite states from the renormalizable Lagrangian of the second site. As a result, the corresponding operators are generated either at loop level, or directly at the scale $\Lambda$, in both cases carrying extra powers of $g_\rho / 4\pi$.
This type of UV selection rules in particular realizes the minimal coupling (MC) condition, as defined in Ref.~\cite{Giudice:2007fh}. The couplings of neutral matter fields to the on-shell gauge bosons $S G_{\mu \nu} G^{\mu \nu}$, $S W_{\mu \nu} W^{\mu \nu}$, $S B_{\mu \nu} B^{\mu \nu}$, $|H|^2 G_{\mu \nu} G^{\mu \nu}$, $|H|^2 \gamma_{\mu \nu} \gamma^{\mu \nu}$, $(D_\mu H)^\dagger \sigma^i (D_\nu H) W^{i \mu \nu}$, and $(D_\mu H)^\dagger (D_\nu H) B^{\mu \nu}$ can not be generated at tree level in $N$-site models and hence carry a $g_\rho^2 / (4 \pi)^2$ suppression. Part of the couplings listed above, namely $S X^2$ and $|H|^2 X^2$, are expected to be loop-level even without MC if $S$ and the Higgs are PNGBs. Notice that, unlike the $1/N$ ``loop" suppression of large-$N$ theories, the MC loop suppression in $N$-site models is automatic rather than optional.

\subsection{Effective Lagrangian construction and choice of the operator basis}
\label{sec:valid}

In the EFT obtained after integrating out the UV degrees of freedom, one generically expects all operators compatible with the gauge and approximate global symmetries and their breaking patterns. We assume that the degrees of freedom of the EFT are chosen such that all the associated symmetries are manifest, and hence our power counting~(\ref{eq:nda}) directly applies to all EFT operators.
It would however be impractical to perform physical analyses with the full set of possible operators, given that some of them are redundant.
Our goal here is to find the minimal set of operators obeying the power counting~(\ref{eq:nda}), to which the full set can be reduced. This task is nontrivial because certain manipulations with the effective operators leading to the reduction of their total number also explicitly break of the power counting.
In other words, if we simply eliminate all the redundant operators without paying attention to the size of the corrections induced to the remaining ones, the resulting operator coefficients may not follow the power counting and the presence of symmetries may become hidden in correlations between different coefficients. This will become clear in the following part of this section where we construct a set of operators capturing the leading interactions of the new spin-zero state $S$ with SM fields. They can be described by operators of dimension five at least (unless $S$ features some additional symmetries which we do not consider) and those are the only ones we will consider.

We will first discuss the case of a scalar $S$. Let us start by analyzing the set of dimension-five operators containing the fields $S$, $H$ as well as two derivatives
\bea\label{eq:derops}
&\displaystyle {\cal O}_1 = {1 \over f} |D_\mu H|^2 S \;\;\;\;\;\;
{\cal O}_2 = {i \over f} (H^\dagger D_\mu H) \partial^\mu S + \text{h.c.}  \;\;\;\;\;\;
{\cal O}_3 = {1\over f} \partial_\mu |H|^2 \partial^\mu S &  \\
&\displaystyle {\cal O}_4 = {1\over f} (H^\dagger \Box H) S  + \text{h.c.} \;\;\;\;\;\;
{\cal O}_5 = {1\over f}  |H|^2 \Box S  \nn
\eea
and discuss which of these operators can be eliminated without breaking our power counting.

{\setlength{\leftmargini}{\descriptionmargin}%
\begin{description}
\item[${\cal O}_3$, ${\cal O}_5$]
Integration by parts relates these two operators which have the same symmetry breaking properties, i.e. invariance under $S \to S+c$ but not under $H\to H+c$. Hence, we can safely eliminate either of them without conflicting with our power counting. We choose to eliminate $\mathcal{O}_3$ in favor of $\mathcal{O}_5$:
\be \label{eq:O3-O5}
{\cal O}_3 \to - {\cal O}_5 \,.
\ee

\item[${\cal O}_1$, ${\cal O}_4$, ${\cal O}_5$]
The importance of the symmetry breaking structure can be seen when considering the operator ${\cal O}_1$, which can be expressed in terms of other operators using integration by parts
\be \label{eq:dhdhs}
{\cal O}_1 \to {1 \over 2} \left({\cal O}_5 - {\cal O}_4  \right)\,.
\ee
One immediately realizes that the two operators in the {\it r.h.s.} break the Higgs shift symmetry, unlike the one they originate from.
This poses a problem if $S$ is a generic composite state because the operator ${\cal O}_1$, not carrying in this case any loop suppression, after using the equality~(\ref{eq:dhdhs}) gives rise to two operators breaking the Higgs shift symmetry with unsuppressed coefficients, in contradiction to our power counting rules~(\ref{eq:nda}). In general these types of problems are also expected to arise in a theory with both $S$ and $H$ being PNGBs, but with a different size of shift symmetry breaking. The fact that $H\to H+c$ breaking is suppressed will now be encoded into a correlation of ${\cal O}_4$ and ${\cal O}_5$ coefficients defined by~\autoref{eq:dhdhs}.

If we proceed further in this direction, the operators ${\cal O}_4$ and ${\cal O}_5$ generated by ${\cal O}_1$ can be eliminated by the field redefinitions
\be \label{eq:shredef}
H \to H\left(1+ {\alpha_H\over f} S\right)\;\;\;,\;\;\;
S\to S +{\alpha_S\over f} |H|^2
\ee
which give the following modifications of the kinetic and mass terms
\bea
\delta {\cal L}_{\text{kin}}^{H} = - {\alpha_H} {\cal O}_4 + \dots \,,\;\; &
\;\; \delta {\cal L}_{\text{mass}}^{H} =  {2\alpha_H \over f} \mu^2 S |H|^2 + \dots \,,\;\; \\
\delta {\cal L}_{\text{kin}}^{S} = - {\alpha_S} {\cal O}_5 + \dots \,,\;\; &
\;\; \delta {\cal L}_{\text{mass}}^{S} = - {\alpha_S \over f} M^2 S |H|^2 + \dots  \,,\;\;
\eea
where $-\mu^2$ is the mass parameter of the Higgs doublet and the ellipses stand for higher-order operators.
Hence, by appropriately choosing $\alpha_S$ and $\alpha_H$, the $\mathcal{O}_4$ and $\mathcal{O}_5$ operators can be eliminated. In return, one receives modifications of the remaining operators, e.g. $S|H|^2$ gets shifted by
\be \label{eq:SHsq}
\left(-\alpha_S {M^2 \over f} + 2\alpha_H {\mu^2 \over f}  \right)  S |H|^2\,,
\ee
and if $S$ in not a PNGB we have $\alpha_S M^2 / f  \lesssim m_\rho^2/f$, which does not feature the loop suppression factor expected for the $H$ shift breaking operators. The Higgs shift symmetry is now hidden in the correlation among the different coefficients, and can not be reconstructed by making order of magnitude power counting estimates.
For instance, if we had performed an analysis in the basis where ${\cal O}_{1,4,5}$ are eliminated and assumed that the operator $S |H|^2$ has a coefficient of the size $\sim (M^2/f)$ without keeping track of all the correlations, we would have obtained an excessive mixing of the $S$ and the physical Higgs boson, which has in fact to be loop suppressed. Consequently, we would have overestimated the impact of $S$ on  Higgs physics.
Had we instead assumed a loop suppressed coefficient for the operator $S |H|^2$, we would then have underestimated the physical effects originally triggered by the ${\cal O}_1$ operator, e.g. $S \to hh$ decay rate.
Similar problems appear with higher-order operators ($S |H|^4, S^2 |H|^2, \dots$) and can be traced back to the field redefinitions~(\ref{eq:shredef}) which cause an unsuppressed explicit breaking of the $H$ shift symmetry when $\alpha_{S,H} \sim 1$.
We hence conclude that the elimination of the operator ${\cal{O}}_{1}$ (as e.g. in Ref.~\cite{Gripaios:2016xuo}) would not allow to apply our power counting to the operator basis if $S$ is a generic scalar. Instead, we use the equality~(\ref{eq:dhdhs}) to rewrite ${\cal O}_5$ in terms of ${\cal O}_4$ and ${\cal O}_1$, while ${\cal O}_4$ can be eliminated using one of the field redefinitions~(\ref{eq:shredef}).

Note that a similar situation occurs in the construction of the SILH basis of the dimension-six operators for the composite Higgs boson. The Higgs shift-symmetry preserving operator $\text{Tr}[D_\mu U (D^\mu U)^\dagger]$ (\ref{eq:kinU}), giving rise to the Higgs kinetic term, also produces two operators $|H|^2 |D_\mu H|^2$ and $\partial_\mu |H|^2 \partial^\mu |H|^2$. Each of the latter breaks $H \to H+c$, but the specific linear combination of them coming from $\text{Tr}[D_\mu U (D^\mu U)^\dagger]$ is shift invariant.
Then, the operator $|H|^2 |D_\mu H|^2$ is removed by an order-one shift-symmetry-breaking field redefinition $H\to H(1+\gamma |H|^2/f)$, while $\partial_\mu |H|^2 \partial^\mu |H|^2$ remains in the SILH basis with an unsuppressed coefficient. At the level of dimension-six operators, this field redefinition does however not generate any SILH power counting breaking, besides the one associated with $\partial_\mu |H|^2 \partial^\mu |H|^2$.
It is also important to notice that, in this case, two operators of the full initial set break the power counting estimates even before any manipulations. We do however not expect a similar situation, contradicting our starting assumption, to occur at the level of dimension-five operators involving $S$.

In case $S$ is a PNGB with the same properties as the Higgs, expressing ${\cal{O}}_{1}$ in terms of ${\cal{O}}_{4,5}$ poses no problem since the former has to be loop suppressed. Afterwards, ${\cal{O}}_{4,5}$ can be eliminated by the redefinitions~(\ref{eq:shredef}) with loop suppressed $\alpha_{S,H}$, without introducing any breaking of the power counting.

\item[${\cal O}_2$]
This operator can safely be removed by gauge field redefinitions and expressed in terms of operators of $\partial_\mu S\: \bar q \gamma^\mu q$ type.
\end{description}%
}

We conclude that one of the operators with derivatives, $|D_\mu H|^2 S$, can not be removed if $S$ and $H$ have different natures, i.e. if $S$ is a generic composite resonance or a PNGB with a larger breaking size than that of the Higgs. Note that one of the field redefinitions~(\ref{eq:shredef}), which is not used in this case, can still serve to eliminate one operator. Let us now discuss operators of the type $S^n |H|^{2m}$.

{\setlength{\leftmargini}{\descriptionmargin}%
\begin{description}
\item[$S^n |H|^{2m}$]
Elimination of operators of this type can also lead to a violation of the power counting rules.
By applying the equations of motion ({\it e.o.m.})\ of the PNGB $S$ or $H$, one unavoidably generates $H$ shift symmetry breaking terms containing derivatives and not carrying loop suppression factors. For example, for the operator $S|H|^2$, we get
\be \label{eq:elimsh2}
{y_t^2 \over 16\pi^2} {m_\rho^2 \over f} S |H|^{2} \; \to \; {1 \over f} S H^\dagger \Box H  \;\; \text{or} \;\;  {1 \over f} \Box S |H|^2.
\ee
This occurs because any product of the $S$ and $H$ fields enters the {\it e.o.m.} with a loop suppression while the kinetic terms of $H$ or $S$ are unsuppressed.
Analogous problems appear if one attempts to eliminate any operator of the type $S^n |H|^{2m}$.

Hence, we are only left with a possibility to use the $S$ {\it e.o.m.} when $S$ is a generic scalar.
Given that the {\it e.o.m.} in that case contains unsuppressed terms of the type $S^n$, it can be used to re-express $S^n |H|^{2m}$ in terms of other operators without absorbing the loop suppression.
However, with a generic $S$, one can run into a different problem related to the necessary tuning of the $S$ mass. Since the coefficient of the operator $S^2$ brings the main contribution to the physical $S$ mass $M$, it has to be tuned down with respect to the power counting estimate in the same way as $M$. For instance, if we apply the $S$ {\it e.o.m.} to the operator $S |H|^{2}$, we obtain
\be
{y_t^2 \over 16\pi^2} {m_\rho^2 \over f} S |H|^{2} \to  {y_t^2 \over 16\pi^2} {1 \over f} {m_\rho^2 \over M^2} {\Box }S |H|^{2} + \dots
\ee
i.e., the resulting operator coefficients are enhanced by the degree of tuning $m_\rho^2/M^2$ of the $S$ mass.\footnote{Notice that analogous enhancements would not occur when applying the PNGB $H$ {\it e.o.m.}. The physical Higgs mass, which has to be tuned (see discussion of \autoref{sec:SUV}), then receives different contributions, e.g. direct UV contributions encoded in the $|H|^2$ operator, but also IR ones arising from loops involving the top Yukawa $y_t \bar q_L H t_R$.  Generically, no tuning of the $|H|^2$ coefficient is expected at the $S$ mass scale.}
The only two operators of $S^n |H|^{2m}$ type which can be eliminated without problems are $S^2 |H|^2$ and $S^3 |H|^2$ because the coefficients of $S^2$ or $S^3$ in the equation of motion of $S$ are expected to be neither tuned, nor loop suppressed. The $S$ field redefinition allowing to eliminate one of these two operators is precisely the one of \autoref{eq:shredef}, which remained unused since the operator $|D_\mu H|^2 S$ can not be excluded for a generic $S$.
\end{description}%
}

We can now complete the discussion of dimension-five operators. The remaining operators of the form $\partial_\mu S \bar q \gamma^\mu q$, $S\partial_\mu S \partial^\mu S $ can be removed by the fermion and $S$ field redefinitions without breaking the power counting. Up to dimension five, the minimal set of operators preserving the power counting in all the discussed scenarios for $S$ can be chosen to be
\begin{center}
\renewcommand{\arraystretch}{1.6}
\begin{tabular}{c | c}
$S X^2  \;\;,\;\; S^{2,4}\;\;$ &  $S  |D_\mu H|^2\;\;,\;\;  S^{3,5}\;\;$ \\
\hline
$S \bar q H q\;\;\;,\;\;\; S^2|H|^2\;\;$ & $\;\;S|H|^2\;\;\;,\;\;\; S|H|^4\;\;\;,\;\;\; S^3 |H|^2$
\end{tabular}
\end{center}
where $X^2$ stands for $X_{\mu \nu}^i X^{i\,\mu \nu}$ or $\frac12 \epsilon_{\mu \nu \rho \sigma}X^{i\,\mu \nu} X^{i\,\rho \sigma}\equiv X_{\mu \nu}^i \tilde X^{i\,\mu \nu}$ with $X=G,W,B$ corresponding to the $SU(3)_c$, $SU(2)_L$ and $U(1)_Y$ gauge field strengths. The presence of the canonically normalized kinetic term for $S$ is understood. For definiteness, we assume that the VEV of $S$ vanishes when $\langle H \rangle=0$.
We also assume CP conservation so that a CP-even $S$ only couples to $X_{\mu \nu}^i X^{i\,\mu \nu}$ and a CP-odd $S$ only to $X_{\mu \nu}^i \tilde X^{i\,\mu \nu}$. The operators in the two upper blocks are invariant under $H\to H+c$ shifts but break $S\to S+c$, while the ones in the lower blocks break both shift symmetries.

Part of the operators above can be removed in specific scenarios. The operators in the left blocks are allowed regardless of $S$ CP properties (up to a change of gauge field strength to its dual), while the right blocks are forbidden by CP conservation if $S$ is a pseudo-scalar. In the case of a CP-even $S$, one redundant operator can still be removed from this set. For a PNGB $S$, one can eliminate $S  |D_\mu H|^2$ without breaking our power counting while, for a generic $S$, one can remove either $S^2 |H|^2$ (as done below) or $S^3 |H|^2$.

As a final remark, it is important to mention that the operations leading to the construction of the given basis do not violate any of the UV selection rules identified for the large-$N$ and multisite models, namely the possible loop suppression of the couplings of $S$ to the Higgs and gauge bosons. Their preservation trivially follows from the fact that all the modifications related to the field redefinitions that were applied to $S$ and $H$ are loop suppressed, while the redefinitions of the gauge bosons and fermions are never relevant in this respect.

\section{Model classification and phenomenology according to the dynamics of the singlet field}
\label{sec:HSscen}

With different assumptions regarding the dynamics of the singlet, we now present a description of several next-to-minimal scenarios for the composite Higgs boson and the singlet field, using the operator basis and power counting rules developed in \autoref{sec:SUV}. Each of the scenarios is intended to capture the main distinct features of well-motivated UV completions in a consistent way.
Among the plethora of possible operators, we focus on those describing the leading interactions of $S$ with SM fields, which are responsible for the $S$ production and decays at collider experiments. For this reason, operators of dimension five at most will be considered.
We will also present the implications for Higgs physics observables at the level of dimension-six operators. Their coefficients are already well constrained by the EW precision tests and the Higgs data.
The lowest dimension of the operators appearing after integrating $S$ out increases by at least one unit as $S$ is the only available dimension-one singlet, while the lowest-dimensional gauge singlet combination of SM fields is $|H|^2$.
This means that, in the dimension-six low-energy Lagrangian for the SM fields, there will be no operators coming from the UV Lagrangian with dimension higher than five. Therefore, for the sake of study of $S$ effects at high and low energies, it is consistent to limit our study to ${\cal{L}}^{\text{UV}}_{\leq5} (S,\text{SM})+{\cal{L}}^{\text{IR}}_{\leq6} (\text{SM})$.

\subsection{Classification}
\label{sec:Sscenario}

We classify the different scenarios according to the nature of $S$. The typical magnitude of the operator coefficients is presented in \autoref{tab:tab1} for each case.

\begin{itemize}

\item{{\it Generic scalar}}

In this case, we assume that $S$ is a scalar particle without any specific feature distinguishing it from typical composite resonances. Hence, the only selection rules affecting the power counting are those following from the shift symmetry of the Higgs field, which is dominantly broken by the top quark Yukawa coupling.
Even though generic scalars are expected to have a mass of the same size as other composite resonances, we assume that $M$ is accidentally lighter than $m_\rho$.

\item{{\it Inert scalar}}

Here we impose an additional constraint with respect to the previous scenario, which is a $N_f g_\rho^2/(4\pi)^2$ suppression of $S$ couplings to gauge bosons or the Higgs boson, dictated by some UV selection rules. We do not fix exactly which of the two types of operators gains the loop factor, but there must be at least one.
The operators already carrying the loop suppressions from the Higgs shift symmetry breaking do not get an additional suppression.
 Since the only difference with respect to the generic case is the aforementioned $N_f g_\rho^2 / (4\pi)^2$ factors in the operators ${\cal O}_{X}$ or ${\cal O}_{H,H1,H2,H3,H4}$ (subscripts correspond to that of the corresponding coefficients in \autoref{tab:tab1}), we do not show explicitly the corresponding coefficients.

\item{{\it PNGB scalar with shift symmetry broken by partial compositeness}}

In this scenario, we assume that $S$ is a scalar arising as a Goldstone boson, similarly to the Higgs doublet, and that its shift symmetry is also broken by PC. Hence, we expect couplings to SM fermions of $y_q S \bar q H q$ type, the largest one being that of the top quark. The rest of the $H$ and $S$ shift symmetry breaking couplings, not involving SM fermions, has to carry a loop suppression factor $3 y_t^2/16 \pi^2$. The same suppression also applies to the estimate for the $S$ mass $M^2 \sim (3 y_t^2 /16 \pi^2) \, m_\rho^2$, making $S$ naturally lighter than other composite states.

\item{{\it Generic or inert pseudo-scalar, PNGB pseudo-scalar with shift symmetry broken by PC}}

These three scenarios can be obtained from the previous ones by assuming that $S$ is now CP-odd. In this case, the couplings of $S^n |H|^m$ type with $n$ odd as well as the $S|D_\mu H|^2$ operator are forbidden. The coefficient of the $S \bar q H q$ operator becomes purely imaginary. Unlike in the generic and inert scalar cases, the operator ${\cal O}_{H4}$ can now not be eliminated.

Finally, the couplings to field strengths $X_{\mu\nu}X^{\mu\nu}$ get substituted by couplings to $X_{\mu\nu}\tilde X^{\mu\nu}$.
We expect that the coupling coefficients remain unchanged, in particular, they should be suppressed in the PNGB case. Contrary to Ref.~\cite{Brivio:2017ije}, we argue that this suppression indeed appears despite the fact that under the shift $S\to S+c$ the operators $S X_{\mu\nu}\tilde X^{\mu\nu}$ change only by a total derivative. Indeed, since $S X_{\mu\nu}\tilde X^{\mu\nu}$ operators contribute to the divergence of the current associated to the symmetry under which $S$ shifts~\cite{Peccei:2006as}, they have to vanish if no explicit or anomalous breaking of the symmetry is present, in order to satisfy the corresponding Ward identity. For instance, the coefficient of the operator coupling the neutral pion to a pair of photons in the chiral Lagrangian is exactly that appearing in the divergence of the axial current computed in the UV.

\begin{table}
\centering
\renewcommand{\arraystretch}{1.6}
\begin{tabular}{ c|c|c|c|c|c|}
\cline{2-6}
&\multicolumn{2}{c|}{scalar} & \multicolumn{3}{c|}{pseudo-scalar}  \\
\cline{2-6}
& generic &  PNGB  & generic & PNGB (PC) & PNGB (anom.)  \\
\hline
\multicolumn{1}{|c|}{$k_X\,SX^2$}
& ${g_X^2 \over g_\rho^2} {1 \over f} $  & $ {3 y_t^2 \over (4 \pi)^2}{g_X^2 \over g_\rho^2} {1 \over f}$  &  $  {g_X^2 \over g_\rho^2} {1 \over f} $
&${3 y_t^2 \over (4 \pi)^2}{g_X^2 \over g_\rho^2}{1 \over f}$
&${N_f^{(X)} g_X^2  \over (4 \pi)^2}{1 \over f}$\\
\hline
\multicolumn{1}{|c|}{$k_q\,S \bar q H q$} & $y_q {1\over f}$ & $y_q {1\over f}$  & $i y_q {1\over f}$ & $i y_q {1\over f}$  &  ---  \\
\hline
\multicolumn{1}{|c|}{$k_{H}\, S  |D_\mu H|^2$} &  ${1\over f}$ & --- & --- & --- & ---   \\
\hline
\multicolumn{1}{|c|}{\begin{tabular}{c}$ k_{H1}\, S |H|^2,$ \\[-3mm] $k_{H2}\, S |H|^4/f^2,$\\[-3mm] $k_{H3}\, S^3 |H|^2/f^2 $\end{tabular}} &  $ \,{3 y_t^2 \over (4 \pi)^2} {m_\rho^2 \over f}$ & ${3 y_t^2 \over (4 \pi)^2} {m_\rho^2 \over f}$  & --- & --- & ---   \\
\hline
\multicolumn{1}{|c|}{$  k_{H4}\,S^2 |H|^2$} &  --- & $ {3 y_t^2 \over (4 \pi)^2} {m_\rho^2\over f^2}$ &   ${3 y_t^2 \over (4 \pi)^2} {m_\rho^2\over f^2}$ & $ {3y_t^2 \over (4 \pi)^2} {m_\rho^2\over f^2}$  & $ {\tilde N_f g_\rho^2 \over (4 \pi)^2 } {3 y_t^2 \over (4 \pi)^2} {m_\rho^2\over f^2}$ \\
\hline
\multicolumn{1}{|c|}{$k_M\,S^2\;,\; k_4\,S^4/f^2$} &  $ m_\rho^2$ & $ {3 y_t^2 \over (4 \pi)^2} m_\rho^2$ &   $m_\rho^2$ & $ {3y_t^2 \over (4 \pi)^2} m_\rho^2$  & $ {\tilde N_f g_\rho^2 \over (4 \pi)^2 } m_\rho^2 $   \\
\hline
\multicolumn{1}{|c|}{$k_3\,S^3 \;,\;k_5\,S^5/f^2$} &  $ {m_\rho^2\over f}$ & $ {3 y_t^2 \over (4 \pi)^2} {m_\rho^2 \over f}$ &  --- &  ---  &  ---  \\
\hline
\end{tabular}
\caption{Estimated size of the dimension-five operators involving $S$ corresponding to the scenarios described in the text. We do not list the operators for the inert scalar and pseudo-scalar since they are trivially obtained from the generic ones multiplying  them by a loop factor $N_f g_\rho^2 / (4 \pi)^2$. $X=G,W,B$ corresponds to $SU(3)_c$, $SU(2)_L$ and $U(1)_Y$ gauge field strengths, $X^2$ stands for either $X_{\mu \nu}^i X^{i\,\mu \nu}$ (for scalar $S$) or $X_{\mu \nu}^i \widetilde X^{i\,\mu \nu}$ (pseudo-scalar $S$) while $g_X$ is a corresponding SM gauge coupling, $\bar q q$ is a bilinear of SM fermions and $y_q$ is a corresponding SM Higgs Yukawa coupling. For PNGB pseudo-scalar with anomaly breaking $N_f^{(X)}$ are the coefficients of anomalies associated to SM fields while $\tilde N_f$ is a number of hyperquarks. The empty entries correspond to the operators which are either redundant or not expected to be generated in a given scenario.
}
\label{tab:tab1}
\end{table}

\newpage
\item{{\it PNGB pseudo-scalar with shift symmetry broken by anomaly}}

In this scenario, the shift symmetry breaking of a pseudo-scalar PNGB $S$ is induced by anomalies associated with the SM and strong sector gauge fields.  The strength of the corresponding anomalous interaction with SM gauge bosons $S X_{\mu \nu} \widetilde X^{\mu \nu}$ is proportional to the effective multiplicity of states $N_f^{(X)}$ generating each type of anomalies. The anomaly associated with the new strong dynamics generates a mass $M \simeq (\tilde N_f^{1/2}/ 4\pi) m_\rho$, with $\tilde N_f$ independent of the SM anomalous couplings coefficients $N_f^{(X)}$. We thus have enough sources of symmetry breaking to generate the $S$ mass and the couplings to SM fields. We will therefore assume that PC couplings do not break the $S$ shift symmetry and do not generate Yukawa-like interactions of $S$ with SM fermions.
One can in principle also consider a variation of this scenario, without the anomaly related to the new strong dynamics, but with an additional PC-induced shift symmetry breaking, giving rise to the interaction $S \bar q H q$.
\end{itemize}

\subsection{Direct searches}

Let us briefly analyze the current status and prospects for the direct detection of the new scalar resonances. Not aiming at a comprehensive study of this subject, we will concentrate on the PNGB scenarios,  in which a mild $S$
mass does not require any further tuning.

\begin{figure}
\centering
\includegraphics[width=.6\textwidth]{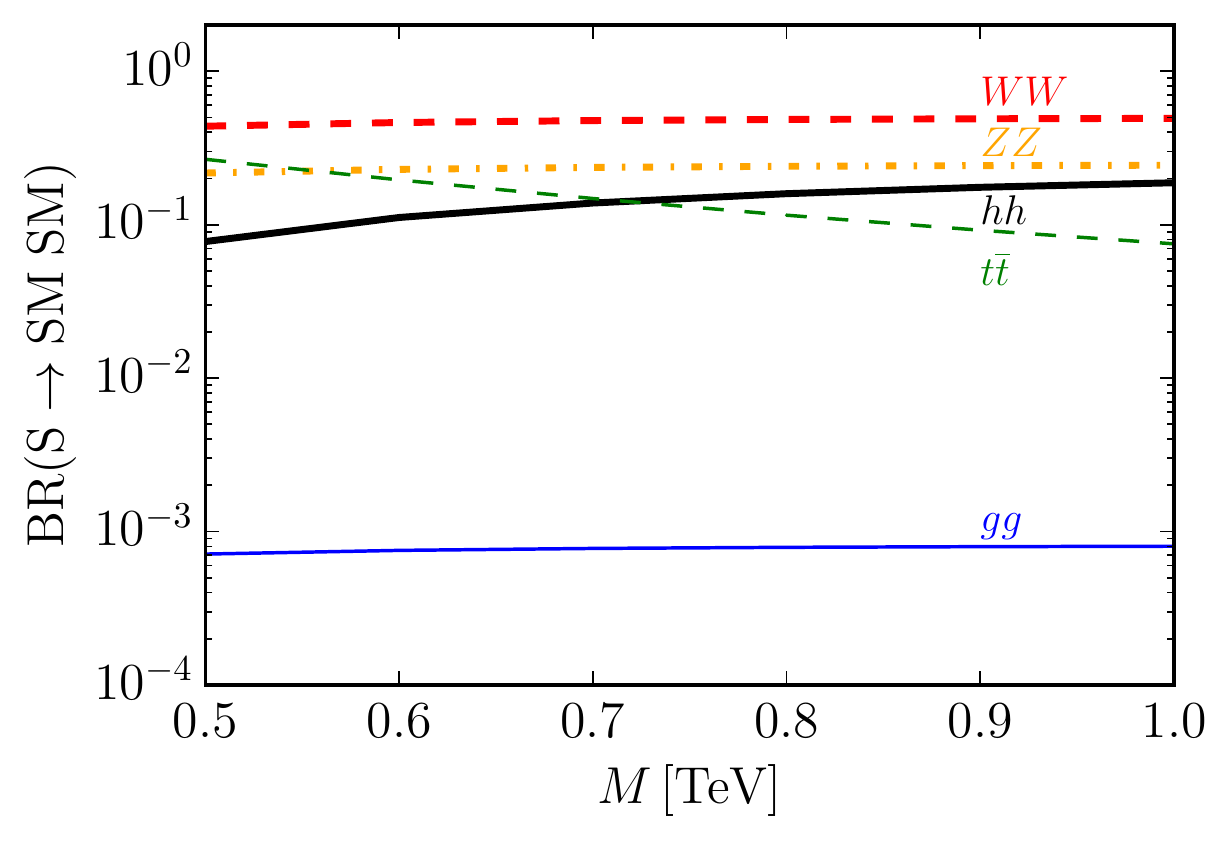}
\caption{Main $S$ branching ratios as a function of the mass for the PNGB scalar (in the pseudo-scalar scenario, $S$ decays almost exclusively into a pair of $t$ quarks). We have assumed $g_\rho = 4\pi$ and all $k_i$ couplings fixed to the unit. Note that $f$ and $m_\rho$ can then be obtained from the expressions in \autoref{tab:tab1}.}
\label{fig:brs}
\end{figure}

The singlet $S$ can be produced at colliders
mainly via gluon fusion. The corresponding cross section at 13\,TeV for $g_\rho
= 4\pi$ ranges from $\sim 1$\,pb for $M = 500$\,GeV, to $\sim 0.04$
pb for $M = 1$\,TeV in the scalar case. These numbers are much smaller in the
pseudo-scalar case, what reflects the fact
that the scalar production cross section is driven by the mixing with the Higgs boson (note,
however, that this behavior can be significantly different if $g_\rho$ is sensibly smaller than $4\pi$).
The main branching ratios of $S$ as functions of its mass, in the
scalar scenario, are shown in \autoref{fig:brs}. All $k_i$ couplings have been
set to the unity.
The large branching ratio into massive gauge bosons, as well as into the Higgs,
is inherited from the sizable mixing with the latter. Provided the $\kappa_X$ couplings remain smaller than $\sim (4\pi)^2 g_\rho^2 \kappa_H$, the decay rates into $WW, ZZ$ and $HH$ at large $M$ are approximately 50\%, 25\% and 25\%, as suggested by the Goldstone-equivalence theorem. On the contrary, in the pseudo-scalar scenario,
$S$ decays almost exclusively into a pair of top quarks, with the second largest decay ratio into a pair of $b$ quarks being of the order of $4\times 10^{-4}$.

Previous studies (see for example Ref.~\cite{Buttazzo:2015bka}) have
estimated the reach of resonant searches in the $HH$ channel at the LHC. The
most optimistic bound ranges from $\sim 0.1$\,pb for $M = 500$\,GeV to $\sim 0.02$
pb for $M = 1$\,TeV. Searches for $ZZ$ are expected to be more constraining, the
bounds being of order $0.04$\,pb for $M = 500$\,GeV and $0.01$\,pb for $M = 1$\,TeV.
Thus, when accounting for the branching ratios depicted in the figure, direct
searches are not very sensitive to the high-mass region. The latter is better tested
by indirect searches, to which we devote the next section.

On the other side, using $2.3$\,fb$^{-1}$ of integrated luminosity
collected at $13$\,TeV, current searches for resonances decaying into pairs of
tops~\cite{CMS-PAS-B2G-15-002} 
bound the pseudo scalar production cross sections
at the $100$\,pb level for $M = 500$\,GeV and at the $3$\,pb level for $M =
1$\,TeV. So, the high-mass region is not expected to be probed in the
near future.

\subsection{Impact of a scalar \texorpdfstring{$S$}{S} on low-energy physics}
\label{sec:hiphy}

In this section,  we examine the impact of the singlet $S$ on Higgs physics and compare it with the generic effects caused by the composite nature of the Higgs boson or due to the heavier strong-sector resonances.
In practice, we look at the dimension-six operators obtained after integrating out the singlet $S$.
First of all, we note that the operators with a pseudo-scalar $S$ do not affect the
low-energy dimension-six operators neither at tree nor at one loop level, hence
we will only consider the scalar $S$ scenarios in the reminder of this section.
As already pointed out, integrating out $S$ at tree level increases operator
dimensions by at least one unit for each external $S$ field, with the minimal
increase corresponding to a substitution $S\to |H|^2$. This means that the
operators ${\cal O}_{H3}$, ${\cal O}_{4}$ and ${\cal O}_{5}$ will not contribute
at dimension-six at all. The operators ${\cal O}_{H2}$, ${\cal O}_{H4}$ and
${\cal O}_{3}$ will only contribute to the $|H|^6$ operator of the Higgs
potential. The form of the resulting corrections to the Higgs potential together
with its derivation is given in \app{sec:intout}, while in this section our main
focus will be on the operators which can affect the Higgs couplings to other SM
fields. Corresponding dimension-six effective operators in the SILH
basis~\cite{Giudice:2007fh} are listed in \autoref{tab:tab2} together with the
estimated size of their coefficients. In addition to the contributions due to
$S$, we present the power counting estimates for the generic contributions of
the strong dynamics which are independent of $S$, and hence relevant for both
scalar and pseudo-scalar $S$ scenarios. Several comments are in order. First, we
will not discuss the full basis, but only those operators which are affected by
$S$. Second, the operator $\alpha |D_\mu H|^2 |H|^2$ was eliminated by a field
redefinition $H\to H (1 - \alpha/2 |H|^2/f^2)$ which also leads to shifts in
${\cal O}_q$ and ${\cal O}_H$. Third, the operator $|H|^2 W_{\mu \nu}^i
W^{i\,\mu \nu}$ was traded for five operators in \autoref{tab:tab2} using the
identity~\cite{Contino:2013kra}
\be \label{eq:opident}
{g^2 \over v^2} |H|^2 W_{\mu \nu}^i W^{i\,\mu \nu} = 4 ({\cal O}_W- {\cal O}_B+ {\cal O}_{HB}- {\cal O}_{HW})+{\cal O}_\gamma \,.
\ee
where in the definition of ${\cal O}_W$ and ${\cal O}_B$ we used $H^\dagger \overleftrightarrow{D}_\mu H = H^\dagger (D_\mu H) - (D_\mu H)^\dagger H$. Hence all five operators generated by $S$ have a coefficient of the same parametric size, which is not the case for the contributions coming from the Higgs compositeness effects. This difference can be understood from the fact that the EW gauge couplings $g$ and $g^\prime$ break the Higgs shift symmetry, hence generically the shift symmetry breaking operators involving $SU(2)_L$ and $U(1)_Y$ gauge bosons, like ${\cal O}_W,{\cal O}_B,{\cal O}_{HB},{\cal O}_{HW}$ do not require loops with $y_t$. This argument does not apply to ${\cal O}_\gamma$ containing a coupling with two photons $|H|^2 \gamma_{\mu \nu} \gamma^{\mu \nu}$, since the external photons can not break the Higgs shift symmetry. Hence an additional loop factor $3 y_t^2/(4\pi)^2$ in ${\cal O}_\gamma$ coefficient.
This explains why in \autoref{tab:tab2} the coefficients of the five operators of \autoref{eq:opident} coming from the generic compositeness effects have different number of $y_t$-loops, while all the operators coming from $S$, instead, have to involve a  $y_t^2/(4\pi)^2$ factor just because it controls the $h-S$ mixing. On top of that, the generic estimate for the size of operators ${\cal O}_{HW}$ and ${\cal O}_{HB}$ contains a loop factor $g_\rho^2 / (4\pi)^2$ which comes from the MC assumption as defined in Ref.~\cite{Giudice:2007fh} (see Ref.~\cite{Liu:2016idz} for a further clarification). In the following, we will analyse both possibilities, with and without (``general SILH" of Ref.~\cite{Liu:2016idz}) MC assumption.

Finally, we mention again that all the estimates for the inert scalar scenario, which are not given explicitly in \autoref{tab:tab2}, can be obtained from those for the generic scalar by multiplying coefficients $k_{X}$ and (or) $k_{H,H1,H2,H3,H4}$ by a factor $N_f g_\rho^2/(4\pi)^2$.

\begin{table}%
\centering%
\scalebox{.75}{%
\renewcommand{\arraystretch}{1.6}%
\begin{tabular}{|c|c|c|c|c|}
\cline{3-5}
\multicolumn{2}{c|}{} &\multicolumn{2}{c|}{effect of scalar $S$} & \multirow{2}{*}{\parbox{2.4cm}{\raggedright compositeness effects [+MC]}} \\
\cline{3-4}
\multicolumn{2}{c|}{} & generic & PNGB  &   \\ \hline
${\cal O}_g$
& ${g_S^2\over  v^2}|H|^2 G_{\mu\nu} G^{\mu\nu}$
& $k_{g}   k_{H1} \,{3 y_t^2 \over (4 \pi)^2}{1 \over g_\rho^2}  {m_\rho^2 \over M^2} \xi$
& $k_{g}   k_{H1} \,{9 y_t^4 \over (4 \pi)^4}{1 \over g_\rho^2}  {m_\rho^2 \over M^2} \xi$
& $ c_g {3y_t^2\over (4\pi)^2}  {1 \over g_\rho^2} \xi$   \\
${\cal O}_\gamma$
& ${g^{\prime 2} \over  v^2}|H|^2 B_{\mu\nu} B^{\mu\nu}$
& $(k_W+k_{B})   k_{H1} \,{3 y_t^2 \over (4 \pi)^2}{1 \over g_\rho^2}  {m_\rho^2 \over M^2} \xi$
& $(k_W+k_{B})   k_{H1} \,{9 y_t^4 \over (4 \pi)^4}{1 \over g_\rho^2}  {m_\rho^2 \over M^2} \xi$
& $ c_\gamma {3y_t^2\over (4\pi)^2}  {1 \over g_\rho^2} \xi$   \\ \hline
${\cal O}_W$
& ${i g\over 2 v^2}(H^\dagger \sigma^i\overleftrightarrow D_\mu H) (D_\nu W^{\mu\nu})^i$
& $ 4 k_{W} k_{H1} \,{3 y_t^2 \over (4 \pi)^2}{1 \over g_\rho^2}  {m_\rho^2 \over M^2} \xi$
& $4 k_{W} k_{H1}\, {9 y_t^4 \over (4 \pi)^4}{1 \over g_\rho^2} {m_\rho^2 \over M^2} \xi$
& $ c_W {1 \over g_\rho^2} \xi$  \\
${\cal O}_B$
& ${i g^\prime\over 2 v^2}(H^\dagger \overleftrightarrow D_\mu H) (\partial_\nu B^{\mu\nu})$
& $ - 4 k_{W} k_{H1} \,{3 y_t^2 \over (4 \pi)^2}{1 \over g_\rho^2}  {m_\rho^2 \over M^2} \xi$
& $-4 k_{W} k_{H1}\, {9 y_t^4 \over (4 \pi)^4}{1 \over g_\rho^2} {m_\rho^2 \over M^2} \xi$
& $ c_B {1 \over g_\rho^2} \xi$  \\
${\cal O}_{HW}$
& ${i g \over v^2}(D_\mu H)^\dagger \sigma^i (D_\nu H)W^{i \mu \nu}$
& $ - 4 k_{W} k_{H1} \,{3 y_t^2 \over (4 \pi)^2}{1 \over g_\rho^2}  {m_\rho^2 \over M^2} \xi$
& $-4 k_{W}  k_{H1}\, {9 y_t^4 \over (4 \pi)^4}{1 \over g_\rho^2} {m_\rho^2 \over M^2} \xi$
&   $c_{HW} {1 \over g_\rho^2} \xi \left[g_\rho^2 \over (4\pi)^2 \right]$  \\
${\cal O}_{HB}$
& ${i g^\prime \over v^2}(D_\mu H)^\dagger (D_\nu H) B^{\mu \nu}$
& $ 4 k_{W} k_{H1} \,{3 y_t^2 \over (4 \pi)^2}{1 \over g_\rho^2}  {m_\rho^2 \over M^2} \xi$
& $4 k_{W}  k_{H1}\, {9 y_t^4 \over (4 \pi)^4}{1 \over g_\rho^2} {m_\rho^2 \over M^2} \xi$
& $c_{HB} {1 \over g_\rho^2} \xi \left[g_\rho^2 \over (4\pi)^2 \right]$  \\ \hline
${\cal O}_q$
& ${1\over v^2} \bar q H q |H|^2$
& $y_q k_{H1} \,  \left(k_q -{k_H \over 2}\right)  {3 y_t^2 \over (4\pi)^2} {m_\rho^2 \over M^2} \xi$
& $y_q k_{H1}  k_q  {3 y_t^2 \over (4\pi)^2} {m_\rho^2 \over M^2} \xi$
& $c_q y_q \xi$  \\ \hline
${\cal O}_H$
& ${1\over 2 v^2} \partial_\mu |H|^2 \partial^\mu |H|^2$
& $ k_{H1} \left( k_{H1} {3 y_t^2 \over (4 \pi)^2} {m_\rho^2 \over M^2} -  k_{H} \right){3 y_t^2 \over (4 \pi)^2}{m_\rho^2 \over M^2} \xi$
& $k_{H1}^2  {9 y_t^4 \over (4 \pi)^4} {m_\rho^4 \over M^4} \xi$
& $c_H \xi$  \\ \hline
\end{tabular}%
}%
\caption{Contributions of a generic or PNGB scalar $S$ to the dimension-six operators in the SILH basis, together with the contributions coming from the generic strong dynamics. Coefficients for the inert scalar can be obtained by multiplying with additional loop factors, as described in the text. $\sigma^i$ are Pauli matrices. $c_i$ are order-one coefficients.   The loop suppression factors in square brackets apply if we impose MC on the Higgs sector. Except for the two operators $\mathcal{O}_W$ and $\mathcal{O}_B$, the effects of $S$ can potentially dominate over the other strong sector effects if $S$ is sufficiently light and $g_\rho$ sufficiently small.
}
\label{tab:tab2}
\end{table}

Now let us turn to the discussion of the phenomenological implications of the operators in \autoref{tab:tab2}. An exhaustive analysis of the generic Higgs compositeness effects was performed in Ref.~\cite{Giudice:2007fh}, hence we limit ourselves to a discussion of the physical effects which can be dominated by the presence of a new resonance $S$. We compare the coefficients assuming $M^2/m_\rho^2 > 3 y_t^2/ (4 \pi)^2$ for a generic scalar and $M^2/m_\rho^2 > (3 y_t^2/ (4 \pi)^2)^2$ for a PNGB $S$, i.e. we require that the tuning of the $S$ mass parameter is not too high. By inspection of the coefficients in \autoref{tab:tab2}, we find six operators whose coefficients can in principle be dominated by the contributions of $S$:  ${\cal O}_g$, ${\cal O}_\gamma$, ${\cal O}_{HW}$, ${\cal O}_{HB}$, ${\cal O}_{q}$, ${\cal O}_{H}$.

\begin{figure}
\centering
\includegraphics[width=.8\textwidth]{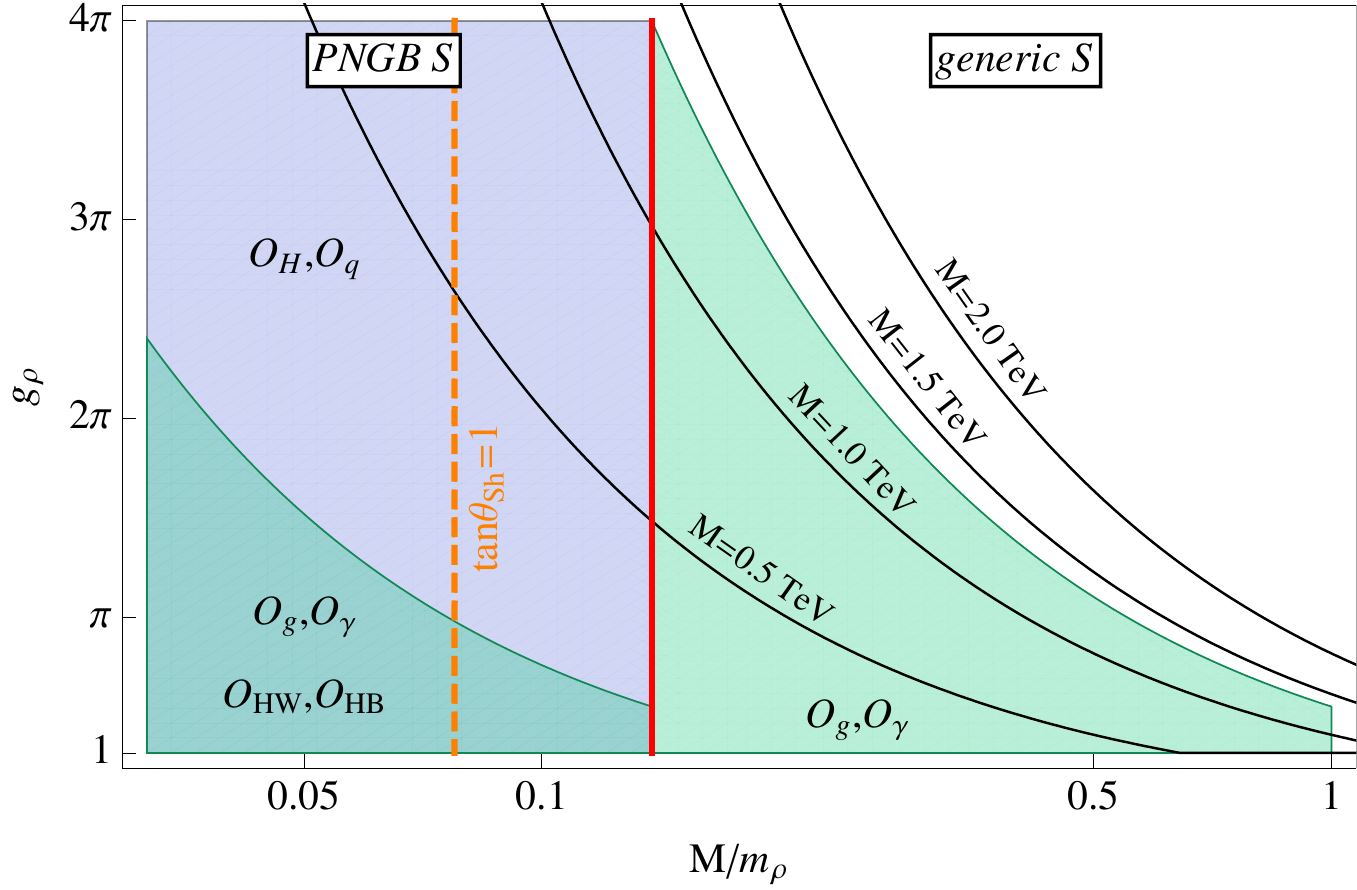}
\caption{In the shaded regions of the $\left({M \over m_\rho}, {g_\rho} \right)$ plane, the contributions to the operators of \autoref{tab:tab2} induced by a PNGB (left from the red line) or a generic $S$ (right from the red line) can become larger than those of the generic strong dynamics and thus alter the dynamics of the Higgs boson. The red line corresponds to $M^2 = {3 y_t^2 \over (4\pi)^2} m_\rho^2$, which we take as a lower bound on the generic $S$ mass and an upper bound on the PNGB $S$ mass. The orange dashed line corresponds to $\tan \theta_{Sh}=1$ (for $\xi=0.1$) and, to the left from this line, the physical observables start being sensitive to interference of multiple $S$-induced operators. The black lines correspond to $\xi = 0.1$, for $M=0.5, \ldots, 2.0$\,TeV ($\xi$ increases towards the upper right corner). Since $\xi = 0.1$ corresponds to the ultimate reach of the LHC with indirect searches for generic compositeness effects only (in the absence of $S$), the black line passing some colored region (i.e. with $S$ effects superseding compositeness effects) means that corresponding indirect $S$ effects are detectable at the LHC.}
\label{fig:ops}
\end{figure}

{\setlength{\leftmargini}{\descriptionmargin}%
\begin{description}
\item[${\cal O}_g$, ${\cal O}_\gamma$]
The contribution of $S$ to these two operators can be dominant when $M^2/m_\rho^2 \lesssim1$ for a generic $S$ (i.e. in the whole regime of validity of our EFT), and when $M^2/m_\rho^2 \lesssim 3 y_t^2/(4\pi)^2$ for a PNGB $S$.
Both conditions require some tuning and correspond to the same suppression of the $S$ mass with respect to the power counting estimates. They can be rewritten as $k_M\lesssim1$ (see \autoref{tab:tab1}).
The main effect of the ${\cal O}_g$, ${\cal O}_\gamma$ operators is a modification of the interaction strength between the Higgs with gluons and photons. But the observables sensitive to these couplings, such as Higgs $\Gamma_{gg}$, $\Gamma_{\gamma \gamma}$ partial widths, also receive sizable contributions from the ${\cal O}_H$ and ${\cal O}_t$ operators. Hence, it is important to check under which conditions the $S$-induced modifications of the Higgs partial widths, coming from the ${\cal O}_g$ and ${\cal O}_\gamma$ operators, can dominate over the generic compositeness contributions induced by ${\cal O}_H$ and ${\cal O}_t$.
The latter induce order-$\xi$ Higgs boson field renormalization and top quark Yukawa coupling distortion. Hence, one can expect that the effective coupling of the Higgs to gluons and photons, mediated by the top loop,
results in a distortion of order $\xi$ compared to the SM expectations:
\be
\delta \mathcal{L} \sim {g_X^2 \over (4 \pi)^2} \xi {h\over v} X^2\,.
\ee
Comparing it to the $S$-mediated direct contribution from \autoref{tab:tab2}, we conclude that the latter can become dominant if $M^2/ m_\rho^2  \lesssim 3 y_t^2/g_\rho^2$ for a generic scalar or if $M^2/ m_\rho^2  \lesssim 9 y_t^4/(4 \pi g_\rho)^2$ for a PNGB $S$, which translates into $k_M\lesssim3 y_t^2/g_\rho^2$ for both cases.

\item[${\cal O}_{HW}$, ${\cal O}_{HB}$]
For a minimally coupled Higgs, one could think that the contribution of $S$ to the coefficients of these two operators becomes comparable or larger than the generic estimates when $M^2/ m_\rho^2  \lesssim 3 y_t^2/g_\rho^2$ for a generic $S$, and $M^2/ m_\rho^2  \lesssim 9 y_t^4/(4 \pi g_\rho)^2$ for a PNGB $S$.
It however seems natural to assume that $S$ is minimally coupled, if the Higgs is. So, in the case of a generic $S$, we should then further suppress our estimates for the $SX^2$ couplings by an additional loop factor: $k_{g,W,B} \to k_{g,W,B}\; g_\rho^2/(4 \pi)^2$ (and recover an inert scalar scenario). No additional factors would be needed for a PNGB $S$ since its LO coupling to gauge bosons is already loop suppressed. After this modification, we see that only the PNGB $S$ can give dominant contributions to ${\cal O}_{HW}$ and ${\cal O}_{HB}$. The main physical process sensitive to these two operators is the $h\to Z\gamma$ decay.

\item[${\cal O}_{q}$] This operator can be sensitive to the presence of a PNGB $S$ in the regime where $M^2/ m_\rho^2  \lesssim 3 y_t^2/(4 \pi)^2$, i.e. for $k_M \lesssim 1$. It affects a variety of Higgs observables, including the partial widths of $h\to qq$, $h\to gg$, $h\to \gamma \gamma$, $h\to Z \gamma$ which all receive contributions from top quark loops.

\item[${\cal O}_{H}$] For this operator also, contributions from $S$ which are larger than that of a generic strong dynamics only arise in the PNGB case, with $k_M \lesssim 1$. Given that this operator leads to a Higgs wave function renormalization, it affects all the Higgs decay and production channels. It also controls the energy growth of the longitudinal gauge bosons scattering amplitudes.
\end{description}%
}

Note that, in all cases, the large modifications of Higgs physics observables come at a price of a large mixing between $h$ and $S$, defined by $\tan {\theta_{hS}} = k_{H1} {3 y_t^2 \over (4\pi)^2} {m_\rho^2 \over M^2} {v \over f}$. When the mixing approaches the order-one level, the interferences between BSM operators become sizable and has to be accounted for when computing their impact on physical observables.
In the generic $S$ scenario, the Higgs-gluon coupling originating from ${\cal O}_g$ can for instance be written as $k_{g} \tan \theta_{hS} {h \over f} G_{\mu \nu}G^{\mu \nu}$ and grows with $\tan \theta_{hS}$. But after having accounted for the $h$ field renormalization induced by ${\cal O}_H \ni \tan \theta_{hS}^2(\partial_\mu h)^2/2$, the $h$-gluons coupling becomes $k_{g} \tan \theta_{hS} (1+\tan \theta_{hS}^2)^{-1/2} {h \over f} G_{\mu \nu}G^{\mu \nu}$, hence one achieves the expected result that the coupling is proportional to $\sin \theta_{hS}$ which saturates to 1.

The effects of $S$ typically lead to order-$\xi$ modifications of the various Higgs observables and distort the pattern of deviations predicted by SILH. %
\autoref{fig:ops} highlights the region of the parameters space where the effects of $S$ dominate over that of the strong sector.

\section{Matching to explicit composite Higgs models}
\label{sec:specmod}

To support the above EFT treatment, we now discuss its matching to two specific CH models.

\subsection{\texorpdfstring{
	$SO(6)\times U(1)^\prime/SO(5)\times U(1)^\prime$
	}{SO(6)xU(1)'/SO(5)xU(1)'}}

Let us first consider a composite Higgs model based on the symmetry breaking pattern $G/H=SO(6)\times U(1)^\prime/SO(5)\times U(1)^\prime$~\cite{Gripaios:2009pe}. The addition of a spectator group $U(1)^\prime$ is required in order for the SM-fermion hypercharges to be correctly reproduced, in the same vein as in the minimal CH model. The ten unbroken and five broken generators, $T$ and $X$ respectively, can be chosen to be
\begin{align}
T^{mn}_{ij} &= -\frac{i}{\sqrt{2}} (\delta^m_i\delta^n_j-\delta^n_i\delta^m_j)~,
 &m<n\in[1,5]~,\\[-1mm]
X^{m6}_{ij} &= -\frac{i}{\sqrt{2}} (\delta^m_i\delta^6_j-\delta^6_i\delta^m_j)~,
 &m\in[1, 5]~.
\end{align}
The SM $SU(2)_L\times U(1)_Y$ gauge group is thus generated by
\begin{align}
 &J_L^1 = \frac{1}{\sqrt{2}}(T^{14}+T^{23})~, &J_L^2 = \frac{1}{\sqrt{2}}(T^{24}-T^{13})~, \\[-1mm]
 &J_L^3 = \frac{1}{\sqrt{2}}(T^{12}+T^{34})~, &J_R^3 = \frac{1}{\sqrt{2}}(T^{12}-T^{34})~,
\end{align}
being the hypercharge defined as $Y = J_R^3 + Y^\prime$ with $Y^\prime$ the generator of $U(1)^\prime$. Under the SM gauge group, the five PNGBs $h_i$ transform as a doublet with hypercharge $Y=1/2$ and a complete singlet. The former can be thus identified with the Higgs degrees of freedom, while the latter gives rise to $S$. The dynamics of the PNGBs is dictated by the Goldstone matrix
\begin{equation}
 U = \exp \left \{ -i \frac{\sqrt{2}}{f} \, h_i \, X^i \right \}~.
\end{equation}
In the unitary gauge, it can conveniently be written as
\begin{equation}
 U = \left(\begin{array}{cccccc}
            \mathbf{1}_{3\times 3} & & & & & \\
            & & & 1-\dfrac{h^2}{f^2+f\sqrt{f^2-h^2-S^2}} & -\dfrac{h S}{f^2+f\sqrt{f^2-h^2-S^2}} & \dfrac{h}{f}\\[5mm]
            & & & -\dfrac{h S}{f^2+f\sqrt{f^2-h^2-S^2}} & 1-\dfrac{S^2}{f^2+f\sqrt{f^2-h^2-S^2}} & \dfrac{S}{f}\\[5mm]
            & & & -\dfrac{h}{f} & -\dfrac{S}{f} & \dfrac{1}{f}\sqrt{f^2-h^2-S^2}
           \end{array}\right)~.
\end{equation}
After having integrated out the heavy states of the second site, the Lagrangian of the model reads
\begin{equation}
 \mathcal{L} = \frac{f^2}{4} \text{Tr} (d_{\mu} d^{\mu}) + \mathcal{L}_\text{Yukawa} - V~,
\end{equation}
where $d_\mu$\footnote{The term $\frac{f^2}{4} \text{Tr} (d_{\mu} d^{\mu})$ can be obtained from the Goldstone bosons kinetic term~(\ref{eq:kinU}) of the two-site model after integrating out the heavy vectorial resonances.} is the projection of the Maurer--Cartan one-form $\omega_\mu = i U^{-1} D_\mu U$ into the broken generators $T^i$, while $\mathcal{L}_\text{Yukawa}$ and $V$ stand for the Yukawa Lagrangian and the potential, respectively. The first term is completely fixed by the coset structure:
\begin{equation}
\frac{f^2}{4} \text{Tr} (d_{\mu} d^{\mu}) = (D_\mu H)^\dagger D^\mu H + \frac{1}{2}(\partial_\mu S)^2 + \frac{1}{2f^2}\bigg[\partial_\mu(H^\dagger H)+\frac{1}{2}\partial_\mu S^2\bigg]^2 + \cdots
\end{equation}
where the ellipsis stands for higher-order terms in the $1/f$ expansion.

$\mathcal{L}_\text{Yukawa}$ and $V$, instead, depend on the elementary-composite fermion mixing~\cite{Panico:2015jxa}. For concreteness, we assume that only the third generation quarks sizably mix with the strong sector, $t_R$ being fully composite while $q_L$ mixes with a single composite resonance transforming in the symmetric representation $\mathbf{20}$ of $SO(6)$. (It turns out that this is the minimal setup for which the leading term in the potential expansion in spurions~\cite{Panico:2015jxa} can lead to EW symmetry breaking.) The left-handed third-generation quarks, $q_L = (t_L, b_L)^T$ can hence be embedded in the following multiplet:
\begin{equation}
  Q_L = \Lambda_L^1 b_L + \Lambda_L^2 t_L  = \frac{1}{2}\left(\begin{array}{cccccccc}
             \multicolumn{4}{c}{\multirow{4}{*}{$\mathbf{0}_{4\times 4}$}}
                    & \zeta b_L & ib_L \\
              & & & & -i \zeta  b_L  & b_L \\
              & & & & \zeta  t_L & it_L \\
              & & & & i\zeta t_L & -t_L \\
             \zeta b_L & -i\zeta b_L & \zeta t_L & i\zeta t_L  &
             \multicolumn{2}{c}{\multirow{2}{*}{$\mathbf{0}_{2\times 2}$}} \\
             ib_L & b_L & it_L & -t_L &  \\
           \end{array}\right)~,
\end{equation}
where $\zeta$ is a free parameter that we take to be real. This makes of $S$ a well-defined CP-odd state.
The Yukawa Lagrangian can then be written (up to order $1/f^2$) as
\begin{multline}
 \mathcal{L}_\text{Yukawa} = \frac{y_t}{\sqrt{2}} \overline{t_R}(U^T)_{6I} (U^T)_{6J} Q_L^{IJ} + \text{h.c.}
\\[-3mm]= -y_t\overline{q_L}\tilde{H} t_R \bigg(1 - \frac{|H|^2}{f^2} - \frac{S^2}{2f^2}\bigg) -iy_t\zeta \frac{S}{f}\bigg(\overline{q_L} \tilde{H} t_R\bigg) + \text{h.c.}
\end{multline}
Besides, the potential can be written as
\begin{align}
V=\:& C_1 f^4\sum_{\alpha} \bigg|(U^T)_{6I} (U^T)_{6J} (\Lambda_L^\alpha)^{IJ}\bigg|^2 + C_2 f^4\sum_{\alpha, a} \bigg|(U^T)_{aI} (U^T)_{6J} (\Lambda_L^\alpha)^{IJ}\bigg|^2
\nn\\=\:& 2C_1\bigg[ f^2 |H|^2 -2 |H|^4 + |H|^2 S^2(\zeta^2-1)\bigg]
\nn\\&+ C_2 \bigg[1+\frac{f^2}{2}|H|^2 (\zeta^2-7) + 4|H|^4 + 2|H|^2S^2(1-\gamma^2) + f^2 S^2(\zeta^2-1)\bigg]~,
\end{align}
$C_1$ and $C_2$ being free dimensionless parameters. These can be traded for the measured values of the Higgs VEV, $v$, and the quartic coupling, $\lambda$. Thus,
\begin{equation}
 V = \mu^2 |H|^2 + \lambda |H|^4 + \lambda f^2(1-2\xi)\bigg(\frac{\zeta^2-1}{\zeta^2-3}\bigg) S^2 + \frac{1}{2}(1-\zeta^2) \lambda S^2 |H|^2.
\end{equation}

$SO(6)$ admits anomalous representations, for it is locally isomorphic to $SU(4)$. They would manifest as a Wess--Zumino--Witten term which, in first approximation, is given by~\cite{Gripaios:2009pe,Gripaios:2016mmi}
\begin{equation}
 \mathcal{L}_{WZW} = \frac{n}{16\pi^2 } \frac{S}{f} \left( g_1^2 B_{\mu\nu} \tilde{B}^{\mu\nu} -  g_2^2 W_{\mu\nu}^i \tilde{W}^{\mu\nu}_i\right)~,
\end{equation}
with $n$ an integer number.
Further subleading contributions from SM fermion loops can generate corrections to this term, as well as $SG^2$ interactions. In summary, leading-order estimates for the coefficients of the relevant operators are:
\begin{center}
\begin{tabular}{ccccccc}\hline\\[-0.2cm]
 $S B^2$ & $S W^2$ & $SG^2$ & $S\overline{q} H q$ & $S|D_\mu H|^2$ & $S|H|^2$ & $S^2$\\[0.2cm]\hline\\[-0.3cm]
 $\dfrac{n g_1^2}{16\pi^2 f}$ & $-\dfrac{n g_2^2}{16\pi^2 f}$ & 0 & $-i\dfrac{y_t \zeta}{f}$  & 0 & 0& $2 \lambda f^2(1-2\xi)\bigg(\dfrac{\zeta^2-1}{\zeta^2-3}\bigg)$\\[0.3cm]\hline
\end{tabular}
\end{center}

The fact that $S|H|^2$ and $S|D_\mu H|^2$ are both vanishing is a consequence of the pseudo-scalar nature of $S$. However, even for a complex $\zeta$, $S|D_\mu H|^2$ vanishes to a first approximation, given that the coset is symmetric and hence no term with an odd number of fields is generated in the sigma model.
Note also that, for $\zeta = 0$, $S$ becomes stable, while $\zeta = 1$ makes $S$ massless for the corresponding Goldstone symmetry remains unbroken. For intermediate values like $\zeta\sim 0.5$ and for $f\sim 2$\,TeV, the singlet mass becomes $M_S \sim 500$\,GeV, in good agreement with our expectations of
equation~\ref{eq:massrelations}.


\subsection{\texorpdfstring{
	$SO(5)\times U(1)_S\times U(1)^\prime/SO(5)\times U(1)^\prime$
	}{SO(5) x U(1)S x U(1)' / SO(5) x U(1)'}}

A different coset also leading to an EW doublet plus a singlet is the  $SO(5)\times U(1)_S\times U(1)^\prime/SO(5)\times U(1)^\prime$, considered in Ref.~\cite{Gripaios:2016mmi}. In this coset, the doublet $h_i$ arises from the $SO(5)/SO(4)$ breaking while the singlet $S$ is associated with the breaking of $U(1)_S$. The coset space $SO(5)/SO(4)$ may be parametrized by the Goldstone matrix
\begin{equation}
  U = \exp \left \{ i \frac{\sqrt{2}}{f} \, h_i \, X^i \right \},
\end{equation}
where $X^{i}$ are the four broken $SO(5)$ generators~\cite{Panico:2015jxa}, while the $S$ dependence is simply given by $\exp(i \sqrt{2} S/f_S )$. We note that, in this case, $S$ has its own decay constant, $f_S$. In unitary gauge, $U$ is given by
\begin{equation}
 U = \left(\begin{array}{ccccc}
            \mathbf{1}_{3\times 3} & & & & \\
            & & & \sqrt{1-\frac{h^2}{f^2}} & \frac{h}{f}\\[2mm]
            & & & -\frac{h}{f} & \sqrt{1-\frac{h^2}{f^2}}
           \end{array}\right)~.
\end{equation}
As before,
\begin{equation}
\frac{f^2}{4} \text{Tr} (d_{\mu} d^{\mu}) +\frac{f_S^2}{4}|\partial_\mu e^{i \frac{\sqrt{2}}{f_S} S }|^2= (D_\mu H)^\dagger D^\mu H  + \frac{1}{2f^2}\partial_\mu(H^\dagger H)^2 +  \frac{1}{2}(\partial_\mu S)^2+\cdots
\end{equation}
In this case, the coset structure does not mix $S$ and $H$. For the matter representations, any consistent embedding of the fermions in $SO(5)$ multiplets with definite charge under $U(1)^\prime$ and $U(1)_S$ may be considered. For simplicity, we restrict ourselves to the case of the MCHM5~\cite{Contino:2006qr,Panico:2011pw}, where each quark is embedded in a $\mathbf{5}$ of $SO(5)$. As before, we consider only the top sector. The embeddings are
\begin{equation}
Q_L=\frac{1}{\sqrt{2}}\left(\begin{array}{c} -b_L \\ i b_L\\ t_L \\ -i t_L \\0 \end{array}\right)_{\frac{2}{3}}^{Z_{Q}},~
T_R=\left(\begin{array}{c} 0 \\ 0\\ 0 \\ 0 \\t_R \end{array}\right)_{\frac{2}{3}}^{Z_{T}},~
\end{equation}
where the subscript refers to the charge under $U(1)^\prime$, while the superscript is the (arbitrary) charge under $U(1)_S$.

The top Yukawa Lagrangian is given to leading order by
\begin{align}
\label{eq:yuk}
 \mathcal{L}_\text{Yukawa} = -y_T \bar{q}_L \tilde{H} t_R \left[1+ i \sqrt{2} \frac{S}{f_S}(Z_Q-Z_T) \right]+\mathrm{h.c.}.
\end{align}
In order to generate a non-trivial potential for $S$, it is necessary to mix the quarks with at least two copies of the composite sector operators, with different charge under $U(1)_S$. The simplest way to do this is to assume that the right-handed top mixes with two different fiveplets of SO(5), $\psi^{(1,2)}_L$, carrying charges $Z_T^{(1)}$ and $Z_T^{(2)}$, but the same $U(1)^\prime$~\cite{Gripaios:2016mmi}. %
 The relevant mixings are then given by
\begin{equation}
\mathcal{L}_{\mathrm{mix}}= \lambda_L  \bar{Q}_L U e^{i \sqrt{2} \frac{S}{f_S}Z_Q} \psi_R+ \lambda_R^{(1)}\bar{T}_R^{(1)} U e^{i \sqrt{2} \frac{S}{f_S} Z_T^{(1)}}   \psi_L^{(1)}+ \lambda_R^{(2)} \bar{T}_R^{(2)} U e^{i \sqrt{2} \frac{S}{f_S}Z_T^{(2)}} \psi_L^{(2)} +\mathrm{h.c.}
\end{equation}
It can be seen that, if one of the $\lambda_R^{(i)}$ vanishes, the dependence on $S$ may be eliminated by choosing the charge of the right-handed top to equal that of the remaining $\psi_L^{(i)}$. For two mixings with operators of different charges, this is no longer possible and the $U(1)_S$ are collectively broken. %
Although interactions involving the light quarks have been omitted, we have implicitly assumed diagonal mixings in order to avoid flavor-violating effects. It should be stressed, however, that contrary to scenarios in which all elementary fields mix with only one composite operator, this is not generally expected from the renormalization group evolution of anarchical couplings in the UV~\cite{Panico:2015jxa}. Instead, further assumptions should be made, such as additional symmetries in the UV, like those proposed in~\cite{Csaki:2008eh}. For example, we could choose flavor universal $Z_Q$ together with $Z_{T}^{(1)} = Z_{T}^{(2)}$. In such a case, as argued above, the $S$ potential would vanish and hence it should come from sources other than partial compositeness, for example anomalies. A different possibility is making $Z_Q$ flavor universal and $\lambda_R^{(1)}\propto\lambda_R^{(2)}$. This second assumption may also be promoted to a discrete symmetry in the strong sector under exchange of the $(1)$ and $(2)$ indices. Under it, we have $\lambda_R^{(1)}=\lambda_R^{(2)}$ (and also other strong sector parameters depending on $(1)$ and $(2)$ should be equal). Of course, this symmetry must be broken by the different $Z_T^{(i)}$, otherwise the $S$ potential is trivial. In this case,  Eq. (\ref{eq:yuk}) is still valid with the replacement $Z_T \rightarrow (Z_T^{(1)}+Z_T^{(2)})/2$. For simplicity, we will assume this symmetry in the table \ref{tab:wilson} below.

Under these assumptions and to leading order in the mixings, the potential has the generic form:
\begin{equation}
V\simeq (\alpha_1+\alpha_2 c_S^{12}) \frac{|H|^2}{f^2}-(\beta_1+\beta_2 c_S^{12}) \frac{|H|^4}{f^4}+\gamma c_S^{12} ,
\end{equation}
where $c_S^{12}=\cos(\sqrt{2}S(Z_T^{(1)}-Z_T^{(2)})/f_S)$. The functions $\alpha_i,~\beta_i,~\gamma$ encode the composite sector resonance contributions to the spectrum and may be straightforwardly computed in a holographic or $N$-site model. Assuming real parameters, the leading contributions to these functions scale with the top mixings as $\alpha_1\sim (\lambda_L)^2/2-((\lambda_R^{(1)})^2+(\lambda_R^{(2)})^2)$, $\alpha_{2}\sim \lambda_R^{(1)}\lambda_R^{(2)}$, $\beta_1\sim (\lambda_L\lambda_R^{(1)})^2+(\lambda_L \lambda_R^{(2)})^2$, $\beta_{2}\sim (\lambda_L)^2 \lambda_R^{(1)}\lambda_R^{(2)}$, $\gamma \sim \lambda_R^{(1)}\lambda_R^{(2)}$.

One may find a non-trivial minimum for
\begin{equation}
c_S^{12}=\pm1,~s_S^{12}=0,~\frac{|H|^2}{f^2}=\xi=\frac{\alpha_1 \pm \alpha_2}{2(\beta_1 \pm \beta_2)},
\end{equation}
given some tuning of the parameters.

Once a tuning is made for the Higgs VEV and mass, the mass of $S$ is given by
\begin{equation}
M_S^2= \frac{\partial^2 V}{\partial S^2}|_{VEV}=\mp 2\frac{(Z_T^{(1)}-Z_T^{(2)})^2}{f_S^2}(\gamma+\alpha_2 \xi-\beta_2 \xi^2),
\end{equation}
which is finite for $\xi \rightarrow 0$ and so leads to a typical mass of $S$ of the size of an untuned PNGB. We also note that, since the potential is even in $S$, no mixing operator $S H^2$ arises, as expected from CP conservation.

In terms of the physical masses and Higgs quartic coupling, we find the potential is given at leading order in $\xi$ by
\begin{align}
V \simeq &~-\mu^2 |H|^2+\lambda |H|^4+(Z_T^{(1)}-Z_T^{(2)})^2 \gamma^\prime f_S^2 S^2+\left(\frac{M_S^2}{2 \mu^2} -\frac{2\gamma^\prime f_S^2 (Z_T^{(1)}-Z_T^{(2)})^2}{ \mu^2}\right)\lambda|H|^2 S^2+\cdots~,
\end{align}
where in order to display explicitly the mass dimensions we rescaled $\gamma \to \gamma^\prime f_S^4$, with $\gamma^\prime$ a dimensionless constant.

For completeness, we note that the quartic interaction of $S$ (in the broken EW phase) is given by $2 M_S^2(Z_T^{(1)}-Z_T^{(2)})^2/f_S^2 S^4$ and gets a contribution from the $|H|^2 S^4$ and $|H|^4 S^4$ higher-dimensional operators not listed above.

Finally, depending only on the quantum numbers of the strong sector, one generically finds an anomalous coupling of $S$ to all the SM gauge bosons, given at leading order by the triangle diagrams with one $U(1)_S$ and two $SU(3)_c$ or $SO(5)$ generators:
\begin{equation}
\mathcal{L}_{WZW}= \frac{1}{16 \pi^2} \frac{S}{f_S} \left(c_3 g_3^2 G_{\mu\nu}^a\tilde{G}^{a~\mu\nu}+c_2 g_2^2  W_{\mu\nu}^i \tilde{W}^{i~\mu\nu}+c_1 g_1^2 B_{\mu\nu}\tilde{B}^{\mu\nu} \right).
\end{equation}
This is in contrast with the $SO(6)/SO(5)$ coset, which lacks an anomalous coupling of $S$ to gluons or photons~\cite{Gripaios:2016mmi}.

In a simple two-site model where the strong sector operators are interpolated by vector-like fermions, one finds the $c_i$ scale as in Ref.~\cite{Franceschini:2015kwy}:
\begin{equation}
\label{eq:anom}
c_3 = {2\over 3} Z_f n_f d_2 I_3 \, , \; c_2 = {2\over 3} Z_f n_f d_3 I_2 \, , \; c_1 = {2\over 3} Z_f n_f d_2 d_3 Y^2 \, ,
\end{equation}
where $Z_f$ is the arbitrary $U(1)_S$ coupling of the vector-like fermions, $n_f$ is a number of composite generations, $d_{2,3}$ and $I_{2,3}$ are dimension and index of the fermions under $SU(2)_L, SU(3)_c$ respectively, and $Y$ is the hypercharge. For a reference multiplet of top partners of MCHM5, with $Z_f=1$, this results in $c_1 \simeq 2, c_2 = 2, c_3 \simeq 6$ per generation.

We may then summarize the relevant operator coefficients in the following table:
\begin{center}\label{tab:wilson}
\begin{tabular}{cccc}\hline\\[-0.2cm]
 $S B^2$ & $S W^2$ & $SG^2$ & $S\overline{q} H q$ \\[0.2cm]\hline\\[-0.3cm]
 $\dfrac{c_1 g_1^2}{16\pi^2 f}$ & $\dfrac{c_2 g_2^2}{16\pi^2 f}$ & $\dfrac{c_3 g_3^2}{16\pi^2 f}$ & $-i\dfrac{\sqrt{2} y_t }{f_S}\left(Z_Q-\frac{(Z_T^{(1)}+Z_T^{(2)})}{2}\right)$ \\[0.4cm]\hline\\[-0.3cm]
 $S|D_\mu H|^2$ & $S^2$ & $S|H|^2$ & $S^2 |H|^2$ \\[0.2cm]\hline\\[-0.3cm]
 0 & $(Z_T^{(1)}-Z_T^{(2)})^2\gamma^\prime f_S^2$&0& $\lambda\left(\frac{M_S^2}{2 \mu^2} -\frac{2(Z_T^{(1)}-Z_T^{(2)})^2\gamma^\prime f_S^2}{ \mu^2}\right)$\\[0.3cm]\hline
\end{tabular}
\end{center}

\section{Conclusions}
\label{sec:conc}

Solutions to the gauge hierarchy problem often predict a set of new physics states not far above the electroweak scale. The first observed such state could most probably be accommodated in a plethora of explicit models. It would thus be desirable to adopt an approach allowing to derive the crucial features of the underlying theory without considering explicit models in all their variety and details. Aiming at capturing a large class of possible TeV-scale completions of the standard model, the composite Higgs scenarios, we addressed the case of a first spin-zero electroweak singlet $S$ discovery. We used an effective field theory in conjunction with the relevant power counting to estimate the magnitude of observable effects. This theoretical bias connects the pattern of potential signals to the structural features of the UV dynamics.

We distinguished scenarios featuring a new, CP-even or -odd, PNGB or generic composite resonance. Our description of several classes of models within a single framework allows to compare them and judge their respective viability. The effects of the rest of the strong sector was captured through two parameters, namely a typical mass and a typical coupling, as well as selection rules. The scaling of operator coefficients with these parameters was encapsulated in the power counting~(\ref{eq:nda}) which the known classes of explicit UV-complete composite Higgs models were shown to reproduce. We construct bases of dimension-five operators at most, compatible with this power counting, for the different hypotheses about the nature of $S$.
In doing so, we identified and addressed several problems related to the elimination of redundant operators in the presence of two coupled spin-zero states among which at least one possesses an approximate shift symmetry. The solutions we discussed may find other applications: in the construction of a basis of higher-dimensional operators, for an opposite mass hierarchy between $H$ and $S$, or for $S$ featuring additional symmetries such as in EFT approaches to a strongly coupled dark matter particle~\cite{Bruggisser:2016ixa, Bruggisser:2016nzw}.

Besides the direct production and decay of $S$, we discussed its hypothetical implications for Higgs observables. They could be sizably altered by a CP-even $S$, because of its mixing with the Higgs. The contributions of $S$ to the Higgs couplings could distort the SILH pattern produced by a generic composite sector when the parameters of the $S$ Lagrangian are non-generic and, in particular, when the $S$ mass deviates from its power-counting estimate. In general, all the $S$ scenarios we considered have important implications for the Higgs sector and its naturalness. How natural the electroweak symmetry breaking appears in each case thus provides an additional tool to assess their plausibility.

\subsection*{Acknowledgments}

We thank Roberto Contino, Giuliano Panico, Alex Pomarol, Arcadi Santamaria and Lewis Tunstall for useful discussions. C.G. is supported by the European Commission through the Marie Curie Career Integration Grant 631962 and by the Helmholtz Association through the recruitment initiative. L.L. acknowledges support by the S\~{a}o Paulo Research Foundation
(FAPESP) under grants 2012/21436-9 and 2015/25393-0. M.C. is partially supported by the Spanish MINECO under grant FPA2014-54459-P and by the Severo Ochoa Excellence Program under grant SEV-2014-0398.

\appendix

\section{Integrating out \texorpdfstring{$S$}{S}}
\label{sec:intout}

In this section we present a procedure to integrate out the scalar state $S$, and provide a resulting scalar potential for the Higgs field.
We will consider the $S$ Lagrangian of the following general form
\be \label{eq:slag1}
{\cal L}_S={1\over 2} (\partial_\mu S)^2 + \mu_1^3 S+ \mu_2^2 S^2 + \mu_3 S^3
\ee
where $\mu_i^n$ can contain constant terms and operators.
Minimizing it with respect to the $S$ field configurations we find
\be \label{eq:seom1}
\mu_1^3+(2\mu_2^2 -  \Box)S + 3 \mu_3 S^2  = 0
\ee
We assume that the expansion of~(\ref{eq:slag1})  in $S$ is around the true vacuum, i.e. $\langle S \rangle =0$, and consequently there is no $S$ tadpole, or, equivalently, no constant term in $\mu_1^3$. Hence the leading contribution to $\mu_1^3$ has to be proportional to the lowest dimension singlet combination of the SM fields -- $|H|^2$, while the other $\mu_i^n$ can contain the constant contributions with the size around $M$ or $m_\rho$. Hence $(\mu_1^3)^{1/3} \ll (\mu_2^2)^{1/2},\mu_3$ and we can search for the solution for $S$ as an expansion in powers of $\mu_1^3$. The lowest order term of this expansion is proportional to $|H|^2$ and hence, aiming  at obtaining a dimension-six Lagrangian after integrating out $S$, it is sufficient to consider the Lagrangian~(\ref{eq:slag1}) containing the terms up to $S^3$. It also follows that it is enough to retain in $S$ the terms up to $(\mu_1^3)^2$.
The considerations above are sufficient to write down the desired solution for $S$
\be \label{eq:ssol}
S =  {1 \over \Box-2\mu_2^2} \mu_1^3 + {3 \over \Box-2\mu_2^2} \left[ \left({1 \over \Box-2\mu_2^2} \mu_1^3\right)^2 \mu_3 \right]
\ee
Now the $\mu_i^n$ coefficients can be matched to the Lagrangian of \autoref{tab:tab1}. Then substituting the expression for $S$~(\ref{eq:ssol}) back into \autoref{eq:slag1} allows to reproduce the dimension-six low-energy operators for the Higgs physics of \autoref{tab:tab2}. In addition we get the following correction to the SM Higgs Lagrangian $\mu^2 |H|^2 - \lambda |H|^4$
\begin{center}
\renewcommand{\arraystretch}{1.6}
\begin{tabular}{ |c|c|c|}
\cline{2-3}
\multicolumn{1}{c|}{} & generic $S$ &  PNGB $S$   \\
\hline
$|H|^4$ & ${k_{H1}^2 \over 2} {3^2 y_t^4 \over (4\pi)^4} {m_\rho^4 \over M^2 f^2} - k_H k_{H1}{3 y_t^2 \over (4\pi)^2} {\mu^2 m_\rho^2 \over M^2 f^2}$ &
${k_{H1}^2 \over 2} {3^2 y_t^4 \over (4\pi)^4} {m_\rho^4 \over M^2 f^2} $\\
\hline
 & $ k_{H1}{3 y_t^2 \over (4\pi)^2} \left[ k_{H2}{3 y_t^2 \over (4\pi)^2} {m_\rho^4 \over f^4 M^2} + 2 \lambda k_H {m_\rho^2 \over f^2 M^2} \right]$ &
$k_{H1}k_{H2}{3^2 y_t^4 \over (4\pi)^4}  {m_\rho^4 \over f^4 M^2}$ \\
$|H|^6$ &$+ {k_H^2 k_{H1}^2 \over 4} {3^2 y_t^4 \over (4\pi)^4}  {\mu^2 m_\rho^4 \over f^4 M^4} $  &
$+k_{H1}^2 k_{H4} {3^3 y_t^6 \over (4\pi)^6}  {m_\rho^6 \over f^4 M^4} $  \\
& $+k_{H1}^3{3^3 y_t^6 \over (4\pi)^6} \left[ k_{3} {m_\rho^8 \over f^4 M^6} - {k_H} {m_\rho^6 \over f^4 M^4} \right]$ &
$+k_{H1}^3 k_{3} {3^4 y_t^8 \over (4\pi)^8}  {m_\rho^8 \over f^4 M^6}$ \\
\hline
\end{tabular}
\end{center}
Notice that the operators $|H|^4$ and $|H|^6$ received corrections after the field redefinition $H\to H (1 - \alpha/2 |H|^2/f^2)$ used to remove the operator $\alpha |D_\mu H|^2 |H|^2$.


\begin{raggedright}
	\bibliographystyle{apsrev4-1}
	\bibliography{CHSM.bib}

\begin{thebibliography}{44}%
\makeatletter
\providecommand \@ifxundefined [1]{%
 \@ifx{#1\undefined}
}%
\providecommand \@ifnum [1]{%
 \ifnum #1\expandafter \@firstoftwo
 \else \expandafter \@secondoftwo
 \fi
}%
\providecommand \@ifx [1]{%
 \ifx #1\expandafter \@firstoftwo
 \else \expandafter \@secondoftwo
 \fi
}%
\providecommand \natexlab [1]{#1}%
\providecommand \enquote  [1]{``#1''}%
\providecommand \bibnamefont  [1]{#1}%
\providecommand \bibfnamefont [1]{#1}%
\providecommand \citenamefont [1]{#1}%
\newcommand \href@noop[0]{\@secondoftwo}%
\providecommand \href[0]{\begingroup \@sanitize@url \@href}%
\providecommand \@href[1]{\@@startlink{#1}\@@href}%
\providecommand \@@href[1]{\endgroup#1\@@endlink}%
\providecommand \@sanitize@url [0]{\catcode `\\12\catcode `\$12\catcode
  `\&12\catcode `\#12\catcode `\^12\catcode `\_12\catcode `\%12\relax}%
\providecommand \@@startlink[1]{}%
\providecommand \@@endlink[0]{}%
\providecommand \url  [0]{\begingroup\@sanitize@url \@url }%
\providecommand \@url [1]{\endgroup\@href {#1}{\urlprefix }}%
\providecommand \urlprefix  [0]{URL }%
\newcommand \Eprint [0]{\href }%
\providecommand \doibase [0]{http://dx.doi.org/}%
\providecommand \selectlanguage [0]{\@gobble}%
\providecommand \bibinfo  [0]{\@secondoftwo}%
\providecommand \bibfield  [0]{\@secondoftwo}%
\providecommand \translation [1]{[#1]}%
\providecommand \BibitemOpen [0]{}%
\providecommand \bibitemStop [0]{}%
\providecommand \bibitemNoStop [0]{.\EOS\space}%
\providecommand \EOS [0]{\spacefactor3000\relax}%
\providecommand \BibitemShut  [1]{\csname bibitem#1\endcsname}%
\let\auto@bib@innerbib\@empty
\bibitem [{\citenamefont {Contino} \emph {et~al.}(2003)\citenamefont {Contino},
  \citenamefont {Nomura}, and \citenamefont {Pomarol}}]{ch}%
  \BibitemOpen
  \bibfield  {author} {\bibinfo {author} {\bibfnamefont {R.}~\bibnamefont
  {Contino}}, \bibinfo {author} {\bibfnamefont {Y.}~\bibnamefont {Nomura}},
  and \bibinfo {author} {\bibfnamefont {A.}~\bibnamefont {Pomarol}},
  }\href{\doibase 10.1016/j.nuclphysb.2003.08.027}{\bibfield  {journal}
  {\bibinfo  {journal} {Nucl. Phys.} }\textbf {\bibinfo {volume} {B671}}
  (\bibinfo {year} {2003}) \bibinfo {pages} {148}}, \Eprint
  {http://arxiv.org/abs/hep-ph/0306259} {arXiv:hep-ph/0306259}\BibitemShut
  {NoStop}%
\bibitem [{\citenamefont {Agashe} \emph {et~al.}(2005)\citenamefont {Agashe},
  \citenamefont {Contino}, and \citenamefont {Pomarol}}]{Agashe:2004rs}%
  \BibitemOpen
  \bibfield  {author} {\bibinfo {author} {\bibfnamefont {K.}~\bibnamefont
  {Agashe}}, \bibinfo {author} {\bibfnamefont {R.}~\bibnamefont {Contino}},
  and \bibinfo {author} {\bibfnamefont {A.}~\bibnamefont {Pomarol}},
  }\href{\doibase 10.1016/j.nuclphysb.2005.04.035}{\bibfield  {journal}
  {\bibinfo  {journal} {Nucl. Phys.} }\textbf {\bibinfo {volume} {B719}}
  (\bibinfo {year} {2005}) \bibinfo {pages} {165}}, \Eprint
  {http://arxiv.org/abs/hep-ph/0412089} {arXiv:hep-ph/0412089}\BibitemShut
  {NoStop}%
\bibitem [{\citenamefont {Contino}(2011)}]{Contino:2010rs}%
  \BibitemOpen
  \bibfield  {author} {\bibinfo {author} {\bibfnamefont {R.}~\bibnamefont
  {Contino}}, }\emph {\bibinfo {title} {{The Higgs as a Composite
  Nambu-Goldstone Boson}}}, in \href{\doibase 10.1142/9789814327183_0005}{\emph
  {\bibinfo {booktitle} {{Physics of the large and the small, TASI 2009,
  proceedings of the Theoretical Advanced Study Institute in Elementary
  Particle Physics, Boulder, Colorado, USA, 1-26 June 2009}}}} (\bibinfo
  {publisher} {World Scientific}, \bibinfo {year} {2011}) pp. \bibinfo {pages}
  {235--306}, \Eprint {http://arxiv.org/abs/1005.4269}
  {arXiv:1005.4269~[hep-ph]}\BibitemShut {NoStop}%
\bibitem [{\citenamefont {Panico} and \citenamefont
  {Wulzer}(2016)}]{Panico:2015jxa}%
  \BibitemOpen
  \bibfield  {author} {\bibinfo {author} {\bibfnamefont {G.}~\bibnamefont
  {Panico}} and \bibinfo {author} {\bibfnamefont {A.}~\bibnamefont {Wulzer}},
  }\href{\doibase 10.1007/978-3-319-22617-0}{\bibfield  {journal} {\bibinfo
  {journal} {Lect. Notes Phys.} }\textbf {\bibinfo {volume} {913}} (\bibinfo
  {year} {2016}) \bibinfo {pages} {pp.1}}, \Eprint
  {http://arxiv.org/abs/1506.01961} {arXiv:1506.01961~[hep-ph]}\BibitemShut
  {NoStop}%
\bibitem [{\citenamefont {Bellazzini} \emph {et~al.}(2014)\citenamefont
  {Bellazzini}, \citenamefont {Csáki}, and \citenamefont
  {Serra}}]{Bellazzini:2014yua}%
  \BibitemOpen
  \bibfield  {author} {\bibinfo {author} {\bibfnamefont {B.}~\bibnamefont
  {Bellazzini}}, \bibinfo {author} {\bibfnamefont {C.}~\bibnamefont {Csáki}},
  and \bibinfo {author} {\bibfnamefont {J.}~\bibnamefont {Serra}},
  }\href{\doibase 10.1140/epjc/s10052-014-2766-x}{\bibfield  {journal}
  {\bibinfo  {journal} {Eur. Phys. J.} }\textbf {\bibinfo {volume} {C74}}
  (\bibinfo {year} {2014}) \bibinfo {pages} {2766}}, \Eprint
  {http://arxiv.org/abs/1401.2457} {arXiv:1401.2457~[hep-ph]}\BibitemShut
  {NoStop}%
\bibitem [{\citenamefont {Giudice} \emph {et~al.}(2007)\citenamefont {Giudice},
  \citenamefont {Grojean}, \citenamefont {Pomarol}, and \citenamefont
  {Rattazzi}}]{Giudice:2007fh}%
  \BibitemOpen
  \bibfield  {author} {\bibinfo {author} {\bibfnamefont {G.~F.} \bibnamefont
  {Giudice}}, \bibinfo {author} {\bibfnamefont {C.}~\bibnamefont {Grojean}},
  \bibinfo {author} {\bibfnamefont {A.}~\bibnamefont {Pomarol}},  and \bibinfo
  {author} {\bibfnamefont {R.}~\bibnamefont {Rattazzi}}, }\href{\doibase
  10.1088/1126-6708/2007/06/045}{\bibfield  {journal} {\bibinfo  {journal}
  {JHEP} }\textbf {\bibinfo {volume} {06}} (\bibinfo {year} {2007}) \bibinfo
  {pages} {045}}, \Eprint {http://arxiv.org/abs/hep-ph/0703164}
  {arXiv:hep-ph/0703164}\BibitemShut {NoStop}%
\bibitem [{\citenamefont {Frigerio} \emph {et~al.}(2012)\citenamefont
  {Frigerio}, \citenamefont {Pomarol}, \citenamefont {Riva}, and \citenamefont
  {Urbano}}]{Frigerio:2012uc}%
  \BibitemOpen
  \bibfield  {author} {\bibinfo {author} {\bibfnamefont {M.}~\bibnamefont
  {Frigerio}}, \bibinfo {author} {\bibfnamefont {A.}~\bibnamefont {Pomarol}},
  \bibinfo {author} {\bibfnamefont {F.}~\bibnamefont {Riva}},  and \bibinfo
  {author} {\bibfnamefont {A.}~\bibnamefont {Urbano}}, }\href{\doibase
  10.1007/JHEP07(2012)015}{\bibfield  {journal} {\bibinfo  {journal} {JHEP}
  }\textbf {\bibinfo {volume} {07}} (\bibinfo {year} {2012}) \bibinfo {pages}
  {015}}, \Eprint {http://arxiv.org/abs/1204.2808}
  {arXiv:1204.2808~[hep-ph]}\BibitemShut {NoStop}%
\bibitem [{\citenamefont {Bruggisser} \emph {et~al.}()\citenamefont
  {Bruggisser}, \citenamefont {Riva}, and \citenamefont
  {Urbano}}]{Bruggisser:2016ixa}%
  \BibitemOpen
  \bibfield  {author} {\bibinfo {author} {\bibfnamefont {S.}~\bibnamefont
  {Bruggisser}}, \bibinfo {author} {\bibfnamefont {F.}~\bibnamefont {Riva}},
  and \bibinfo {author} {\bibfnamefont {A.}~\bibnamefont {Urbano}},
  }\href@noop{}{ }\Eprint {http://arxiv.org/abs/1607.02474}
  {arXiv:1607.02474~[hep-ph]}\BibitemShut {NoStop}%
\bibitem [{\citenamefont {Espinosa} \emph {et~al.}(2012)\citenamefont
  {Espinosa}, \citenamefont {Gripaios}, \citenamefont {Konstandin}, and
  \citenamefont {Riva}}]{Espinosa:2011eu}%
  \BibitemOpen
  \bibfield  {author} {\bibinfo {author} {\bibfnamefont {J.~R.} \bibnamefont
  {Espinosa}}, \bibinfo {author} {\bibfnamefont {B.}~\bibnamefont {Gripaios}},
  \bibinfo {author} {\bibfnamefont {T.}~\bibnamefont {Konstandin}},  and
  \bibinfo {author} {\bibfnamefont {F.}~\bibnamefont {Riva}}, }\href{\doibase
  10.1088/1475-7516/2012/01/012}{\bibfield  {journal} {\bibinfo  {journal}
  {JCAP} }\textbf {\bibinfo {volume} {1201}} (\bibinfo {year} {2012}) \bibinfo
  {pages} {012}}, \Eprint {http://arxiv.org/abs/1110.2876}
  {arXiv:1110.2876~[hep-ph]}\BibitemShut {NoStop}%
\bibitem [{\citenamefont {von Harling} and \citenamefont
  {Servant}(2016)}]{vonHarling:2016vhf}%
  \BibitemOpen
  \bibfield  {author} {\bibinfo {author} {\bibfnamefont {B.}~\bibnamefont {von
  Harling}} and \bibinfo {author} {\bibfnamefont {G.}~\bibnamefont {Servant}},
  }\href@noop{}{  (\bibinfo {year} {2016})}, \Eprint
  {http://arxiv.org/abs/1612.02447} {arXiv:1612.02447~[hep-ph]}\BibitemShut
  {NoStop}%
\bibitem [{\citenamefont {Gripaios} \emph {et~al.}(2009)\citenamefont
  {Gripaios}, \citenamefont {Pomarol}, \citenamefont {Riva}, and \citenamefont
  {Serra}}]{Gripaios:2009pe}%
  \BibitemOpen
  \bibfield  {author} {\bibinfo {author} {\bibfnamefont {B.}~\bibnamefont
  {Gripaios}}, \bibinfo {author} {\bibfnamefont {A.}~\bibnamefont {Pomarol}},
  \bibinfo {author} {\bibfnamefont {F.}~\bibnamefont {Riva}},  and \bibinfo
  {author} {\bibfnamefont {J.}~\bibnamefont {Serra}}, }\href{\doibase
  10.1088/1126-6708/2009/04/070}{\bibfield  {journal} {\bibinfo  {journal}
  {JHEP} }\textbf {\bibinfo {volume} {04}} (\bibinfo {year} {2009}) \bibinfo
  {pages} {070}}, \Eprint {http://arxiv.org/abs/0902.1483}
  {arXiv:0902.1483~[hep-ph]}\BibitemShut {NoStop}%
\bibitem [{\citenamefont {Belyaev} \emph {et~al.}(2016)\citenamefont {Belyaev},
  \citenamefont {Cacciapaglia}, \citenamefont {Cai}, \citenamefont {Flacke},
  \citenamefont {Parolini}, and \citenamefont {Serôdio}}]{Belyaev:2015hgo}%
  \BibitemOpen
  \bibfield  {author} {\bibinfo {author} {\bibfnamefont {A.}~\bibnamefont
  {Belyaev}}, \bibinfo {author} {\bibfnamefont {G.}~\bibnamefont
  {Cacciapaglia}}, \bibinfo {author} {\bibfnamefont {H.}~\bibnamefont {Cai}},
  \bibinfo {author} {\bibfnamefont {T.}~\bibnamefont {Flacke}}, \bibinfo
  {author} {\bibfnamefont {A.}~\bibnamefont {Parolini}},  and \bibinfo {author}
  {\bibfnamefont {H.}~\bibnamefont {Serôdio}}, }\href{\doibase
  10.1103/PhysRevD.94.015004}{\bibfield  {journal} {\bibinfo  {journal} {Phys.
  Rev.} }\textbf {\bibinfo {volume} {D94}} (\bibinfo {year} {2016}) \bibinfo
  {pages} {015004}}, \Eprint {http://arxiv.org/abs/1512.07242}
  {arXiv:1512.07242~[hep-ph]}\BibitemShut {NoStop}%
\bibitem [{\citenamefont {Liu} \emph {et~al.}(2016)\citenamefont {Liu},
  \citenamefont {Pomarol}, \citenamefont {Rattazzi}, and \citenamefont
  {Riva}}]{Liu:2016idz}%
  \BibitemOpen
  \bibfield  {author} {\bibinfo {author} {\bibfnamefont {D.}~\bibnamefont
  {Liu}}, \bibinfo {author} {\bibfnamefont {A.}~\bibnamefont {Pomarol}},
  \bibinfo {author} {\bibfnamefont {R.}~\bibnamefont {Rattazzi}},  and \bibinfo
  {author} {\bibfnamefont {F.}~\bibnamefont {Riva}}, }\href{\doibase
  10.1007/JHEP11(2016)141}{\bibfield  {journal} {\bibinfo  {journal} {JHEP}
  }\textbf {\bibinfo {volume} {11}} (\bibinfo {year} {2016}) \bibinfo {pages}
  {141}}, \Eprint {http://arxiv.org/abs/1603.03064}
  {arXiv:1603.03064~[hep-ph]}\BibitemShut {NoStop}%
\bibitem [{\citenamefont {Matsedonskyi}(2015)}]{Matsedonskyi:2014iha}%
  \BibitemOpen
  \bibfield  {author} {\bibinfo {author} {\bibfnamefont {O.}~\bibnamefont
  {Matsedonskyi}}, }\href{\doibase 10.1007/JHEP02(2015)154}{\bibfield
  {journal} {\bibinfo  {journal} {JHEP} }\textbf {\bibinfo {volume} {02}}
  (\bibinfo {year} {2015}) \bibinfo {pages} {154}}, \Eprint
  {http://arxiv.org/abs/1411.4638} {arXiv:1411.4638~[hep-ph]}\BibitemShut
  {NoStop}%
\bibitem [{\citenamefont {Cacciapaglia} \emph {et~al.}(2015)\citenamefont
  {Cacciapaglia}, \citenamefont {Cai}, \citenamefont {Flacke}, \citenamefont
  {Lee}, \citenamefont {Parolini}, and \citenamefont
  {Serodio}}]{Cacciapaglia:2015dsa}%
  \BibitemOpen
  \bibfield  {author} {\bibinfo {author} {\bibfnamefont {G.}~\bibnamefont
  {Cacciapaglia}}, \bibinfo {author} {\bibfnamefont {H.}~\bibnamefont {Cai}},
  \bibinfo {author} {\bibfnamefont {T.}~\bibnamefont {Flacke}}, \bibinfo
  {author} {\bibfnamefont {S.~J.} \bibnamefont {Lee}}, \bibinfo {author}
  {\bibfnamefont {A.}~\bibnamefont {Parolini}},  and \bibinfo {author}
  {\bibfnamefont {H.}~\bibnamefont {Serodio}}, }\href{\doibase
  10.1007/JHEP06(2015)085}{\bibfield  {journal} {\bibinfo  {journal} {JHEP}
  }\textbf {\bibinfo {volume} {06}} (\bibinfo {year} {2015}) \bibinfo {pages}
  {085}}, \Eprint {http://arxiv.org/abs/1501.03818}
  {arXiv:1501.03818~[hep-ph]}\BibitemShut {NoStop}%
\bibitem [{\citenamefont {Panico} and \citenamefont
  {Pomarol}(2016)}]{Panico:2016ull}%
  \BibitemOpen
  \bibfield  {author} {\bibinfo {author} {\bibfnamefont {G.}~\bibnamefont
  {Panico}} and \bibinfo {author} {\bibfnamefont {A.}~\bibnamefont {Pomarol}},
  }\href{\doibase 10.1007/JHEP07(2016)097}{\bibfield  {journal} {\bibinfo
  {journal} {JHEP} }\textbf {\bibinfo {volume} {07}} (\bibinfo {year} {2016})
  \bibinfo {pages} {097}}, \Eprint {http://arxiv.org/abs/1603.06609}
  {arXiv:1603.06609~[hep-ph]}\BibitemShut {NoStop}%
\bibitem [{\citenamefont {Galloway} \emph {et~al.}(2010)\citenamefont
  {Galloway}, \citenamefont {Evans}, \citenamefont {Luty}, and \citenamefont
  {Tacchi}}]{Galloway:2010bp}%
  \BibitemOpen
  \bibfield  {author} {\bibinfo {author} {\bibfnamefont {J.}~\bibnamefont
  {Galloway}}, \bibinfo {author} {\bibfnamefont {J.~A.} \bibnamefont {Evans}},
  \bibinfo {author} {\bibfnamefont {M.~A.} \bibnamefont {Luty}},  and \bibinfo
  {author} {\bibfnamefont {R.~A.} \bibnamefont {Tacchi}}, }\href{\doibase
  10.1007/JHEP10(2010)086}{\bibfield  {journal} {\bibinfo  {journal} {JHEP}
  }\textbf {\bibinfo {volume} {10}} (\bibinfo {year} {2010}) \bibinfo {pages}
  {086}}, \Eprint {http://arxiv.org/abs/1001.1361}
  {arXiv:1001.1361~[hep-ph]}\BibitemShut {NoStop}%
\bibitem [{\citenamefont {Panico} \emph {et~al.}(2013)\citenamefont {Panico},
  \citenamefont {Redi}, \citenamefont {Tesi}, and \citenamefont
  {Wulzer}}]{Panico:2012uw}%
  \BibitemOpen
  \bibfield  {author} {\bibinfo {author} {\bibfnamefont {G.}~\bibnamefont
  {Panico}}, \bibinfo {author} {\bibfnamefont {M.}~\bibnamefont {Redi}},
  \bibinfo {author} {\bibfnamefont {A.}~\bibnamefont {Tesi}},  and \bibinfo
  {author} {\bibfnamefont {A.}~\bibnamefont {Wulzer}}, }\href{\doibase
  10.1007/JHEP03(2013)051}{\bibfield  {journal} {\bibinfo  {journal} {JHEP}
  }\textbf {\bibinfo {volume} {03}} (\bibinfo {year} {2013}) \bibinfo {pages}
  {051}}, \Eprint {http://arxiv.org/abs/1210.7114}
  {arXiv:1210.7114~[hep-ph]}\BibitemShut {NoStop}%
\bibitem [{\citenamefont {D'Ambrosio} \emph {et~al.}(2002)\citenamefont
  {D'Ambrosio}, \citenamefont {Giudice}, \citenamefont {Isidori}, and
  \citenamefont {Strumia}}]{DAmbrosio:2002vsn}%
  \BibitemOpen
  \bibfield  {author} {\bibinfo {author} {\bibfnamefont {G.}~\bibnamefont
  {D'Ambrosio}}, \bibinfo {author} {\bibfnamefont {G.~F.} \bibnamefont
  {Giudice}}, \bibinfo {author} {\bibfnamefont {G.}~\bibnamefont {Isidori}},
  and \bibinfo {author} {\bibfnamefont {A.}~\bibnamefont {Strumia}},
  }\href{\doibase 10.1016/S0550-3213(02)00836-2}{\bibfield  {journal} {\bibinfo
   {journal} {Nucl. Phys.} }\textbf {\bibinfo {volume} {B645}} (\bibinfo {year}
  {2002}) \bibinfo {pages} {155}}, \Eprint
  {http://arxiv.org/abs/hep-ph/0207036} {arXiv:hep-ph/0207036}\BibitemShut
  {NoStop}%
\bibitem [{\citenamefont {Azatov} and \citenamefont
  {Galloway}(2012)}]{Azatov:2011qy}%
  \BibitemOpen
  \bibfield  {author} {\bibinfo {author} {\bibfnamefont {A.}~\bibnamefont
  {Azatov}} and \bibinfo {author} {\bibfnamefont {J.}~\bibnamefont {Galloway}},
  }\href{\doibase 10.1103/PhysRevD.85.055013}{\bibfield  {journal} {\bibinfo
  {journal} {Phys. Rev.} }\textbf {\bibinfo {volume} {D85}} (\bibinfo {year}
  {2012}) \bibinfo {pages} {055013}}, \Eprint {http://arxiv.org/abs/1110.5646}
  {arXiv:1110.5646~[hep-ph]}\BibitemShut {NoStop}%
\bibitem [{\citenamefont {Contino} \emph {et~al.}(2013)\citenamefont {Contino},
  \citenamefont {Ghezzi}, \citenamefont {Grojean}, \citenamefont {Muhlleitner},
  and \citenamefont {Spira}}]{Contino:2013kra}%
  \BibitemOpen
  \bibfield  {author} {\bibinfo {author} {\bibfnamefont {R.}~\bibnamefont
  {Contino}}, \bibinfo {author} {\bibfnamefont {M.}~\bibnamefont {Ghezzi}},
  \bibinfo {author} {\bibfnamefont {C.}~\bibnamefont {Grojean}}, \bibinfo
  {author} {\bibfnamefont {M.}~\bibnamefont {Muhlleitner}},  and \bibinfo
  {author} {\bibfnamefont {M.}~\bibnamefont {Spira}}, }\href{\doibase
  10.1007/JHEP07(2013)035}{\bibfield  {journal} {\bibinfo  {journal} {JHEP}
  }\textbf {\bibinfo {volume} {07}} (\bibinfo {year} {2013}) \bibinfo {pages}
  {035}}, \Eprint {http://arxiv.org/abs/1303.3876}
  {arXiv:1303.3876~[hep-ph]}\BibitemShut {NoStop}%
\bibitem [{\citenamefont {Grojean} \emph {et~al.}(2013)\citenamefont {Grojean},
  \citenamefont {Matsedonskyi}, and \citenamefont {Panico}}]{Grojean:2013qca}%
  \BibitemOpen
  \bibfield  {author} {\bibinfo {author} {\bibfnamefont {C.}~\bibnamefont
  {Grojean}}, \bibinfo {author} {\bibfnamefont {O.}~\bibnamefont
  {Matsedonskyi}},  and \bibinfo {author} {\bibfnamefont {G.}~\bibnamefont
  {Panico}}, }\href{\doibase 10.1007/JHEP10(2013)160}{\bibfield  {journal}
  {\bibinfo  {journal} {JHEP} }\textbf {\bibinfo {volume} {10}} (\bibinfo
  {year} {2013}) \bibinfo {pages} {160}}, \Eprint
  {http://arxiv.org/abs/1306.4655} {arXiv:1306.4655~[hep-ph]}\BibitemShut
  {NoStop}%
\bibitem [{\citenamefont {Matsedonskyi} \emph {et~al.}(2016)\citenamefont
  {Matsedonskyi}, \citenamefont {Panico}, and \citenamefont
  {Wulzer}}]{Matsedonskyi:2015dns}%
  \BibitemOpen
  \bibfield  {author} {\bibinfo {author} {\bibfnamefont {O.}~\bibnamefont
  {Matsedonskyi}}, \bibinfo {author} {\bibfnamefont {G.}~\bibnamefont
  {Panico}},  and \bibinfo {author} {\bibfnamefont {A.}~\bibnamefont {Wulzer}},
  }\href{\doibase 10.1007/JHEP04(2016)003}{\bibfield  {journal} {\bibinfo
  {journal} {JHEP} }\textbf {\bibinfo {volume} {04}} (\bibinfo {year} {2016})
  \bibinfo {pages} {003}}, \Eprint {http://arxiv.org/abs/1512.04356}
  {arXiv:1512.04356~[hep-ph]}\BibitemShut {NoStop}%
\bibitem [{\citenamefont {Bellazzini} \emph {et~al.}(2016)\citenamefont
  {Bellazzini}, \citenamefont {Franceschini}, \citenamefont {Sala}, and
  \citenamefont {Serra}}]{Bellazzini:2015nxw}%
  \BibitemOpen
  \bibfield  {author} {\bibinfo {author} {\bibfnamefont {B.}~\bibnamefont
  {Bellazzini}}, \bibinfo {author} {\bibfnamefont {R.}~\bibnamefont
  {Franceschini}}, \bibinfo {author} {\bibfnamefont {F.}~\bibnamefont {Sala}},
  and \bibinfo {author} {\bibfnamefont {J.}~\bibnamefont {Serra}},
  }\href{\doibase 10.1007/JHEP04(2016)072}{\bibfield  {journal} {\bibinfo
  {journal} {JHEP} }\textbf {\bibinfo {volume} {04}} (\bibinfo {year} {2016})
  \bibinfo {pages} {072}}, \Eprint {http://arxiv.org/abs/1512.05330}
  {arXiv:1512.05330~[hep-ph]}\BibitemShut {NoStop}%
\bibitem [{\citenamefont {Veneziano}(1979)}]{eta}%
  \BibitemOpen
  \bibfield  {author} {\bibinfo {author} {\bibfnamefont {G.}~\bibnamefont
  {Veneziano}}, }\href{\doibase 10.1016/0550-3213(79)90332-8}{\bibfield
  {journal} {\bibinfo  {journal} {Nucl. Phys.} }\textbf {\bibinfo {volume}
  {B159}} (\bibinfo {year} {1979}) \bibinfo {pages} {213}}\BibitemShut
  {NoStop}%
\bibitem [{\citenamefont {Panico} and \citenamefont
  {Wulzer}(2011)}]{Panico:2011pw}%
  \BibitemOpen
  \bibfield  {author} {\bibinfo {author} {\bibfnamefont {G.}~\bibnamefont
  {Panico}} and \bibinfo {author} {\bibfnamefont {A.}~\bibnamefont {Wulzer}},
  }\href{\doibase 10.1007/JHEP09(2011)135}{\bibfield  {journal} {\bibinfo
  {journal} {JHEP} }\textbf {\bibinfo {volume} {09}} (\bibinfo {year} {2011})
  \bibinfo {pages} {135}}, \Eprint {http://arxiv.org/abs/1106.2719}
  {arXiv:1106.2719~[hep-ph]}\BibitemShut {NoStop}%
\bibitem [{\citenamefont {'t~Hooft}(1974{\natexlab{a}})}]{'tHooft:1973jz}%
  \BibitemOpen
  \bibfield  {author} {\bibinfo {author} {\bibfnamefont {G.}~\bibnamefont
  {'t~Hooft}}, }\href{\doibase 10.1016/0550-3213(74)90154-0}{\bibfield
  {journal} {\bibinfo  {journal} {Nucl. Phys.} }\textbf {\bibinfo {volume}
  {B72}} (\bibinfo {year} {1974}{\natexlab{a}}) \bibinfo {pages}
  {461}}\BibitemShut {NoStop}%
\bibitem [{\citenamefont {'t~Hooft}(1974{\natexlab{b}})}]{'tHooft:1974hx}%
  \BibitemOpen
  \bibfield  {author} {\bibinfo {author} {\bibfnamefont {G.}~\bibnamefont
  {'t~Hooft}}, }\href{\doibase 10.1016/0550-3213(74)90088-1}{\bibfield
  {journal} {\bibinfo  {journal} {Nucl. Phys.} }\textbf {\bibinfo {volume}
  {B75}} (\bibinfo {year} {1974}{\natexlab{b}}) \bibinfo {pages}
  {461}}\BibitemShut {NoStop}%
\bibitem [{\citenamefont {Witten}(1979)}]{Witten:1979kh}%
  \BibitemOpen
  \bibfield  {author} {\bibinfo {author} {\bibfnamefont {E.}~\bibnamefont
  {Witten}}, }\href{\doibase 10.1016/0550-3213(79)90232-3}{\bibfield  {journal}
  {\bibinfo  {journal} {Nucl. Phys.} }\textbf {\bibinfo {volume} {B160}}
  (\bibinfo {year} {1979}) \bibinfo {pages} {57}}\BibitemShut {NoStop}%
\bibitem [{\citenamefont {Luty}(1995)}]{Luty:1994ua}%
  \BibitemOpen
  \bibfield  {author} {\bibinfo {author} {\bibfnamefont {M.~A.} \bibnamefont
  {Luty}}, }\href{\doibase 10.1103/PhysRevD.51.2322}{\bibfield  {journal}
  {\bibinfo  {journal} {Phys. Rev.} }\textbf {\bibinfo {volume} {D51}}
  (\bibinfo {year} {1995}) \bibinfo {pages} {2322}}, \Eprint
  {http://arxiv.org/abs/hep-ph/9405271} {arXiv:hep-ph/9405271}\BibitemShut
  {NoStop}%
\bibitem [{\citenamefont {Foadi} \emph {et~al.}(2010)\citenamefont {Foadi},
  \citenamefont {Laverty}, \citenamefont {Schmidt}, and \citenamefont
  {Yu}}]{Foadi:2010bu}%
  \BibitemOpen
  \bibfield  {author} {\bibinfo {author} {\bibfnamefont {R.}~\bibnamefont
  {Foadi}}, \bibinfo {author} {\bibfnamefont {J.~T.} \bibnamefont {Laverty}},
  \bibinfo {author} {\bibfnamefont {C.~R.} \bibnamefont {Schmidt}},  and
  \bibinfo {author} {\bibfnamefont {J.-H.} \bibnamefont {Yu}}, }\href{\doibase
  10.1007/JHEP06(2010)026}{\bibfield  {journal} {\bibinfo  {journal} {JHEP}
  }\textbf {\bibinfo {volume} {06}} (\bibinfo {year} {2010}) \bibinfo {pages}
  {026}}, \Eprint {http://arxiv.org/abs/1001.0584}
  {arXiv:1001.0584~[hep-ph]}\BibitemShut {NoStop}%
\bibitem [{\citenamefont {De~Curtis} \emph {et~al.}(2012)\citenamefont
  {De~Curtis}, \citenamefont {Redi}, and \citenamefont
  {Tesi}}]{DeCurtis:2011yx}%
  \BibitemOpen
  \bibfield  {author} {\bibinfo {author} {\bibfnamefont {S.}~\bibnamefont
  {De~Curtis}}, \bibinfo {author} {\bibfnamefont {M.}~\bibnamefont {Redi}},
  and \bibinfo {author} {\bibfnamefont {A.}~\bibnamefont {Tesi}},
  }\href{\doibase 10.1007/JHEP04(2012)042}{\bibfield  {journal} {\bibinfo
  {journal} {JHEP} }\textbf {\bibinfo {volume} {04}} (\bibinfo {year} {2012})
  \bibinfo {pages} {042}}, \Eprint {http://arxiv.org/abs/1110.1613}
  {arXiv:1110.1613~[hep-ph]}\BibitemShut {NoStop}%
\bibitem [{\citenamefont {Arkani-Hamed} \emph {et~al.}(2001)\citenamefont
  {Arkani-Hamed}, \citenamefont {Cohen}, and \citenamefont
  {Georgi}}]{ArkaniHamed:2001nc}%
  \BibitemOpen
  \bibfield  {author} {\bibinfo {author} {\bibfnamefont {N.}~\bibnamefont
  {Arkani-Hamed}}, \bibinfo {author} {\bibfnamefont {A.~G.} \bibnamefont
  {Cohen}},  and \bibinfo {author} {\bibfnamefont {H.}~\bibnamefont {Georgi}},
  }\href{\doibase 10.1016/S0370-2693(01)00741-9}{\bibfield  {journal} {\bibinfo
   {journal} {Phys. Lett.} }\textbf {\bibinfo {volume} {B513}} (\bibinfo {year}
  {2001}) \bibinfo {pages} {232}}, \Eprint
  {http://arxiv.org/abs/hep-ph/0105239} {arXiv:hep-ph/0105239}\BibitemShut
  {NoStop}%
\bibitem [{\citenamefont {Manohar} and \citenamefont
  {Georgi}(1984)}]{Manohar:1983md}%
  \BibitemOpen
  \bibfield  {author} {\bibinfo {author} {\bibfnamefont {A.}~\bibnamefont
  {Manohar}} and \bibinfo {author} {\bibfnamefont {H.}~\bibnamefont {Georgi}},
  }\href{\doibase 10.1016/0550-3213(84)90231-1}{\bibfield  {journal} {\bibinfo
  {journal} {Nucl. Phys.} }\textbf {\bibinfo {volume} {B234}} (\bibinfo {year}
  {1984}) \bibinfo {pages} {189}}\BibitemShut {NoStop}%
\bibitem [{\citenamefont {Gripaios} and \citenamefont
  {Sutherland}(2016)}]{Gripaios:2016xuo}%
  \BibitemOpen
  \bibfield  {author} {\bibinfo {author} {\bibfnamefont {B.}~\bibnamefont
  {Gripaios}} and \bibinfo {author} {\bibfnamefont {D.}~\bibnamefont
  {Sutherland}}, }\href{\doibase 10.1007/JHEP08(2016)103}{\bibfield  {journal}
  {\bibinfo  {journal} {JHEP} }\textbf {\bibinfo {volume} {08}} (\bibinfo
  {year} {2016}) \bibinfo {pages} {103}}, \Eprint
  {http://arxiv.org/abs/1604.07365} {arXiv:1604.07365~[hep-ph]}\BibitemShut
  {NoStop}%
\bibitem [{\citenamefont {Brivio} \emph {et~al.}(2017)\citenamefont {Brivio},
  \citenamefont {Gavela}, \citenamefont {Merlo}, \citenamefont {Mimasu},
  \citenamefont {No}, \citenamefont {del Rey}, and \citenamefont
  {Sanz}}]{Brivio:2017ije}%
  \BibitemOpen
  \bibfield  {author} {\bibinfo {author} {\bibfnamefont {I.}~\bibnamefont
  {Brivio}}, \bibinfo {author} {\bibfnamefont {M.~B.} \bibnamefont {Gavela}},
  \bibinfo {author} {\bibfnamefont {L.}~\bibnamefont {Merlo}}, \bibinfo
  {author} {\bibfnamefont {K.}~\bibnamefont {Mimasu}}, \bibinfo {author}
  {\bibfnamefont {J.~M.} \bibnamefont {No}}, \bibinfo {author} {\bibfnamefont
  {R.}~\bibnamefont {del Rey}},  and \bibinfo {author} {\bibfnamefont
  {V.}~\bibnamefont {Sanz}}, }\href@noop{}{  (\bibinfo {year} {2017})}, \Eprint
  {http://arxiv.org/abs/1701.05379} {arXiv:1701.05379~[hep-ph]}\BibitemShut
  {NoStop}%
\bibitem [{\citenamefont {Peccei}(2008)}]{Peccei:2006as}%
  \BibitemOpen
  \bibfield  {author} {\bibinfo {author} {\bibfnamefont {R.~D.} \bibnamefont
  {Peccei}}, }\bibfield  {booktitle} {\emph {\bibinfo {booktitle} {{Axions:
  Theory, cosmology, and experimental searches. Proceedings, 1st Joint
  ILIAS-CERN-CAST axion training, Geneva, Switzerland, November 30-December 2,
  2005}}}, }\href{\doibase 10.1007/978-3-540-73518-2_1}{\bibfield  {journal}
  {\bibinfo  {journal} {Lect. Notes Phys.} }\textbf {\bibinfo {volume} {741}}
  (\bibinfo {year} {2008}) \bibinfo {pages} {3}}, \bibinfo {note} {[,3(2006)]},
  \Eprint {http://arxiv.org/abs/hep-ph/0607268}
  {arXiv:hep-ph/0607268}\BibitemShut {NoStop}%
\bibitem [{\citenamefont {Buttazzo} \emph {et~al.}(2015)\citenamefont
  {Buttazzo}, \citenamefont {Sala}, and \citenamefont
  {Tesi}}]{Buttazzo:2015bka}%
  \BibitemOpen
  \bibfield  {author} {\bibinfo {author} {\bibfnamefont {D.}~\bibnamefont
  {Buttazzo}}, \bibinfo {author} {\bibfnamefont {F.}~\bibnamefont {Sala}},  and
  \bibinfo {author} {\bibfnamefont {A.}~\bibnamefont {Tesi}}, }\href{\doibase
  10.1007/JHEP11(2015)158}{\bibfield  {journal} {\bibinfo  {journal} {JHEP}
  }\textbf {\bibinfo {volume} {11}} (\bibinfo {year} {2015}) \bibinfo {pages}
  {158}}, \Eprint {http://arxiv.org/abs/1505.05488}
  {arXiv:1505.05488~[hep-ph]}\BibitemShut {NoStop}%
\bibitem [{CMS(2016)}]{CMS-PAS-B2G-15-002}%
  \BibitemOpen
  \href{http://cds.cern.ch/record/2138345}{\emph {\bibinfo {title} {{Search for
  $\mathrm{t\bar{t}}$ resonances in boosted semileptonic final states in pp
  collisions at $\sqrt{s}=13~\mathrm{TeV}$}}}}, \bibinfo {type} {Tech. Rep.}
  \bibinfo {number} {CMS-PAS-B2G-15-002} (\bibinfo  {institution} {CERN},
  \bibinfo {address} {Geneva}, \bibinfo {year} {2016})\BibitemShut {NoStop}%
\bibitem [{\citenamefont {Gripaios} \emph {et~al.}(2017)\citenamefont
  {Gripaios}, \citenamefont {Nardecchia}, and \citenamefont
  {You}}]{Gripaios:2016mmi}%
  \BibitemOpen
  \bibfield  {author} {\bibinfo {author} {\bibfnamefont {B.}~\bibnamefont
  {Gripaios}}, \bibinfo {author} {\bibfnamefont {M.}~\bibnamefont
  {Nardecchia}},  and \bibinfo {author} {\bibfnamefont {T.}~\bibnamefont
  {You}}, }\href{\doibase 10.1140/epjc/s10052-017-4603-5}{\bibfield  {journal}
  {\bibinfo  {journal} {Eur. Phys. J.} }\textbf {\bibinfo {volume} {C77}}
  (\bibinfo {year} {2017}) \bibinfo {pages} {28}}, \Eprint
  {http://arxiv.org/abs/1605.09647} {arXiv:1605.09647~[hep-ph]}\BibitemShut
  {NoStop}%
\bibitem [{\citenamefont {Contino} \emph {et~al.}(2007)\citenamefont {Contino},
  \citenamefont {Da~Rold}, and \citenamefont {Pomarol}}]{Contino:2006qr}%
  \BibitemOpen
  \bibfield  {author} {\bibinfo {author} {\bibfnamefont {R.}~\bibnamefont
  {Contino}}, \bibinfo {author} {\bibfnamefont {L.}~\bibnamefont {Da~Rold}},
  and \bibinfo {author} {\bibfnamefont {A.}~\bibnamefont {Pomarol}},
  }\href{\doibase 10.1103/PhysRevD.75.055014}{\bibfield  {journal} {\bibinfo
  {journal} {Phys. Rev.} }\textbf {\bibinfo {volume} {D75}} (\bibinfo {year}
  {2007}) \bibinfo {pages} {055014}}, \Eprint
  {http://arxiv.org/abs/hep-ph/0612048} {arXiv:hep-ph/0612048}\BibitemShut
  {NoStop}%
\bibitem [{\citenamefont {Csaki} \emph {et~al.}(2009)\citenamefont {Csaki},
  \citenamefont {Falkowski}, and \citenamefont {Weiler}}]{Csaki:2008eh}%
  \BibitemOpen
  \bibfield  {author} {\bibinfo {author} {\bibfnamefont {C.}~\bibnamefont
  {Csaki}}, \bibinfo {author} {\bibfnamefont {A.}~\bibnamefont {Falkowski}},
  and \bibinfo {author} {\bibfnamefont {A.}~\bibnamefont {Weiler}},
  }\href{\doibase 10.1103/PhysRevD.80.016001}{\bibfield  {journal} {\bibinfo
  {journal} {Phys. Rev.} }\textbf {\bibinfo {volume} {D80}} (\bibinfo {year}
  {2009}) \bibinfo {pages} {016001}}, \Eprint {http://arxiv.org/abs/0806.3757}
  {arXiv:0806.3757~[hep-ph]}\BibitemShut {NoStop}%
\bibitem [{\citenamefont {Franceschini} \emph {et~al.}(2016)\citenamefont
  {Franceschini}, \citenamefont {Giudice}, \citenamefont {Kamenik},
  \citenamefont {McCullough}, \citenamefont {Pomarol}, \citenamefont
  {Rattazzi}, \citenamefont {Redi}, \citenamefont {Riva}, \citenamefont
  {Strumia}, and \citenamefont {Torre}}]{Franceschini:2015kwy}%
  \BibitemOpen
  \bibfield  {author} {\bibinfo {author} {\bibfnamefont {R.}~\bibnamefont
  {Franceschini}}, \bibinfo {author} {\bibfnamefont {G.~F.} \bibnamefont
  {Giudice}}, \bibinfo {author} {\bibfnamefont {J.~F.} \bibnamefont {Kamenik}},
  \bibinfo {author} {\bibfnamefont {M.}~\bibnamefont {McCullough}}, \bibinfo
  {author} {\bibfnamefont {A.}~\bibnamefont {Pomarol}}, \bibinfo {author}
  {\bibfnamefont {R.}~\bibnamefont {Rattazzi}}, \bibinfo {author}
  {\bibfnamefont {M.}~\bibnamefont {Redi}}, \bibinfo {author} {\bibfnamefont
  {F.}~\bibnamefont {Riva}}, \bibinfo {author} {\bibfnamefont {A.}~\bibnamefont
  {Strumia}},  and \bibinfo {author} {\bibfnamefont {R.}~\bibnamefont {Torre}},
  }\href{\doibase 10.1007/JHEP03(2016)144}{\bibfield  {journal} {\bibinfo
  {journal} {JHEP} }\textbf {\bibinfo {volume} {03}} (\bibinfo {year} {2016})
  \bibinfo {pages} {144}}, \Eprint {http://arxiv.org/abs/1512.04933}
  {arXiv:1512.04933~[hep-ph]}\BibitemShut {NoStop}%
\bibitem [{\citenamefont {Bruggisser} \emph {et~al.}(2016)\citenamefont
  {Bruggisser}, \citenamefont {Riva}, and \citenamefont
  {Urbano}}]{Bruggisser:2016nzw}%
  \BibitemOpen
  \bibfield  {author} {\bibinfo {author} {\bibfnamefont {S.}~\bibnamefont
  {Bruggisser}}, \bibinfo {author} {\bibfnamefont {F.}~\bibnamefont {Riva}},
  and \bibinfo {author} {\bibfnamefont {A.}~\bibnamefont {Urbano}},
  }\href{\doibase 10.1007/JHEP11(2016)069}{\bibfield  {journal} {\bibinfo
  {journal} {JHEP} }\textbf {\bibinfo {volume} {11}} (\bibinfo {year} {2016})
  \bibinfo {pages} {069}}, \Eprint {http://arxiv.org/abs/1607.02475}
  {arXiv:1607.02475~[hep-ph]}\BibitemShut {NoStop}%
\end{thebibliography}%
\end{raggedright}

\end{document}